\documentclass[ALICE,manyauthors]{cernphprep}

\usepackage[comma,square,numbers,sort&compress]{natbib}
\usepackage{hyperref}
\usepackage{lineno}
\usepackage{multirow}
\usepackage{soul}

\newcommand{\pp}{\ensuremath{\mathrm {p\kern-0.05em p}}}
\newcommand{\PbPb}{\ensuremath{\mbox{Pb--Pb}}}

\newcommand{\sqrtSnn}{\ensuremath{\sqrt{s_{\mathrm{NN}}}}}

\newcommand{\pt}{\ensuremath{p_{\mathrm{T}}}}
\newcommand{\ptch}{\ensuremath{p_{\mathrm{T, ch jet}}}}

\newcommand{\MeVc}{\ensuremath{\mathrm{MeV}\kern-0.05em/\kern-0.02em c}}
\newcommand{\GeVc}{\ensuremath{\mathrm{GeV}\kern-0.05em/\kern-0.02em c}}
\newcommand{\GeVcSq}{\ensuremath{\mathrm{GeV}\kern-0.05em/\kern-0.02em c^2}}

\newcommand{\jpsi}{\ensuremath{{\rm J}\kern-0.02em/\kern-0.05em\psi}}

\newcommand{\ptjet}{\ensuremath{p_{\mathrm{T}}^{\rm ch jet}}}

\newcommand{\ptraw}{\ensuremath{\ptch^{\rm raw}}}
\newcommand{\ptrec}{\ensuremath{\ptch^{\rm rec}}}
\newcommand{\ptgen}{\ensuremath{\ptch^{\rm gen}}}

\newcommand{\kt}{\ensuremath{k_{\mathrm{T}}}}
\newcommand{\vj}{\ensuremath{v_{2}^{\mathrm{ch~jet}}}}
\newcommand{\vjsix}{\ensuremath{v_{6}^{\mathrm{ch~jet}}}}
\newcommand{\vjfull}{\ensuremath{v_{2}^{\mathrm{calo~jet}}}}
\newcommand{\vjjust}{\ensuremath{v_{2}^{\mathrm{jet}}}}
\newcommand{\vpart}{\ensuremath{v_{2}^{\mathrm{part}}}}
\newcommand{\dpt}{\ensuremath{\delta \kern-0.15em p_{\mathrm{T}}}}
\newcommand{\rholocal}{\ensuremath{\rho_{\mathrm{ch~local}}}}
\newcommand{\rhovar}{\ensuremath{\rho_{\mathrm{ch}}(\varphi)}}
\newcommand{\rhoav}{\ensuremath{\langle \rho_{\rm ch} \rangle}}

\newcommand{\ep}{\ensuremath{\Psi_{\mathrm{EP,~2}}}}
\newcommand{\epthree}{\ensuremath{\Psi_{\mathrm{EP,~3}}}}

\newcommand{\epn}{\ensuremath{\Psi_{\mathrm{EP},~n}}}

\newcommand{\rpn}{\ensuremath{\Psi_n}}
\newcommand{\rptwo}{\ensuremath{\Psi_2}}
\newcommand{\phijet}{\ensuremath{\varphi_{\mathrm{jet}}}}

\newcommand{\nin}{\ensuremath{N_{\mathrm{in}}}}
\newcommand{\nout}{\ensuremath{N_{\mathrm{out}}}}

\begin{document}

\begin{titlepage}
    \PHyear{2015}
    \PHnumber{258}      
    \PHdate{16 September}  
    %

    \title{Azimuthal anisotropy of charged jet production \\ in $\mathbf{\sqrt{{\textit s}_{\rm NN}}}$~=~2.76~TeV~Pb--Pb collisions}
    \ShortTitle{Azimuthal anisotropy of charged jet production }   

    \Collaboration{ALICE Collaboration\thanks{See Appendix~\ref{app:collab} for the list of collaboration members}}
    \ShortAuthor{ALICE Collaboration} 

    \begin{abstract}
        We present measurements of the azimuthal dependence of charged jet production in central and semi-central $\sqrt{s_{\mathrm{NN}}}$ = 2.76 TeV Pb--Pb collisions with respect to the second harmonic event plane, quantified as $v_{2}^{\mathrm{ch~jet}}$. Jet finding is performed employing the anti-$k_{\mathrm{T}}$ algorithm with a resolution parameter $R$ = 0.2 using charged tracks from the ALICE tracking system. The contribution of the azimuthal anisotropy of the underlying event is taken into account event-by-event. The remaining (statistical) region-to-region fluctuations are removed on an ensemble basis by unfolding the jet spectra for different event plane orientations independently. Significant non-zero $v_{2}^{\mathrm{ch~jet}}$ is observed in semi-central collisions (30--50\% centrality) for 20 $<$ $p_{\mathrm{T}}^{\rm ch~jet}$ $<$ 90 ${\mathrm{GeV}\kern-0.05em/\kern-0.02em c}$. The azimuthal dependence of the charged jet production is similar to the dependence observed for jets comprising both charged and neutral fragments, and compatible with measurements of the $v_2$ of single charged particles at high $p_{\mathrm{T}}$. Good agreement between the data and predictions from JEWEL, an event generator simulating parton shower evolution in the presence of a dense QCD medium, is found in semi-central collisions. 
    \end{abstract}
\end{titlepage}
\setcounter{page}{2}

\section{Introduction}
The aim of the heavy-ion program at the LHC is to study strongly interacting matter in ultra-relativistic nuclear collisions where the formation of a quark-gluon plasma (QGP), a deconfined state of quarks and gluons, is expected \cite{qgp}. Hard partons that propagate through the collision medium lose energy via (multiple) scattering and gluon radiation \cite{gluonradiation,gluonradiation2}. Jet measurements are used to experimentally explore parton energy loss in the hot and dense medium. Studies at the LHC and RHIC have shown that jet and high-\pt{} single particle production in heavy-ion collisions are suppressed with respect to the expected production in a superposition of independent pp collisions \cite{charged_jets_raa,jets_raa,longStar,longPhenix,Arsene:2003yk,Back:2003qr,Aamodt:2010jd,Aamodt:2011vg,ATLAS:2012dna,CMS:2012aa}. This observation is consistent with energy loss, which is further supported by measurements of dijet energy asymmetry and di-hadron angular correlations \cite{Aad:2010bu,Chatrchyan:2012nia,Aad:2012vca}. 

In non-central Pb--Pb collisions, the initial overlap region of the colliding nuclei projected into the plane perpendicular to the beam direction has an approximately elliptic shape. Jets emitted along the minor axis of the ellipse (defined as the \emph{in-plane} direction) on average traverse less medium - and are therefore expected to lose less energy - than jets that are emitted along the major axis of the ellipse (the \emph{out-of-plane} direction). The dependence of jet production on the angle relative to the second-harmonic symmetry plane $\Psi_2$ (the symmetry plane angles \rpn{} define the orientations of the symmetry axes of the initial nucleon distribution of the collision) can be used to probe the path-length dependence of jet energy loss. This dependence is quantified by the parameter \vj{}, the coefficient of the second term in a Fourier expansion of the azimuthal distribution of jets relative to symmetry planes $\Psi_n$,
\begin{equation}\label{eq:flowdud}
    \frac{\mathrm{d}N}{\mathrm{d}\left(\phijet - \rpn\right)} \propto 1 + \sum_{n=1}^{\infty} 2 v_n^{\mathrm{jet}} \cos\left[n\left(\phijet - \rpn\right)\right],
    \end{equation}
    where \phijet~denotes the azimuthal angle of the jet.
    
    In central collisions, the average distance that a jet propagates through the medium is approximately equal in the in-plane and out-of-plane directions, therefore a small \vj~is expected. In semi-central collisions the average in-medium distance is shorter, while the relative difference between the average distances in-plane and out-of-plane is larger, hence a non-zero \vj~is expected. Fluctuations in the initial distribution of nucleons within the overlap region can lead to additional contributions to \vj{} and higher harmonic coefficients in the Fourier decomposition. 

    The path-length dependence of parton energy loss is of particular interest because it is sensitive to the underlying energy-loss mechanism. For collisional (elastic) energy loss, the amount of lost energy depends linearly on path length, while for radiative (inelastic) energy loss, the dependence is quadratic due to interference effects \cite{lsquare,lsquare2}. Some strong-interaction models based on the AdS/CFT correspondence suggest an even stronger path-length dependence \cite{dominguez,marquet}. Earlier studies of the $v_2$ of high-\pt{} single particles have already tested the path-length dependence of energy loss \cite{highpt,highptv2CMS,Adams:2004wz,2013wop,oslo}. Comparisons of these results to theoretical calculations have shown that the $v_2$ is sensitive to several aspects of the medium evolution, including the effects of longitudinal and transverse expansion and the life time of the system until freeze-out \cite{Renk:2010qx}. It is therefore important to measure multiple observables that are sensitive to the path-length dependence of energy loss, such as recoil yields of charged particles and jets \cite{Adams:2006yt,Aamodt:2011vg,Adam:2015doa}. Jets are expected to better represent the original parton kinematics and provide more detailed information on energy loss. Theoretical predictions from JEWEL, which couples parton shower evolution to the presence of a QCD medium with a density derived from Glauber simulations \cite{jewel1,jewel2}, have shown that a finite \vjjust{} is expected for non-central collisions at the LHC. Similar results have been found in \vjjust{} studies in heavy-ion collisions generated by the AMPT model \cite{ampt,zapp2}. A first measurement of \vjfull{} of jets comprising both charged and neutral fragments has been reported by the ATLAS collaboration \cite{atlas}. The results presented in this paper extend the \vj{} measurement to a lower \pt{} range (\pt{} $>$ 30 \GeVc{} for central collisions and \pt{} $>$ 20 \GeVc{} for semi-central collisions).

    In this article, measurements of \vj{} of $R = 0.2$ charged jets reconstructed with the anti-\kt{} jet finder algorithm in \PbPb{} collisions with  0--5\% and 30--50\% collision centrality are presented. The largest experimental challenge in jet analyses in heavy-ion collisions is the separation of the jet signal from the background of mostly low-\pt{} particles from the underlying event and from unrelated scatterings that take place in the collision. The jet energy is corrected on a jet-by-jet basis using an estimate of the background transverse momentum density which takes into account the dominant flow harmonics $v_2$ and $v_3$ of the background event-by-event, as will be described in Sections \ref{sec:ep} and \ref{sec:21}. The coefficient \vj{} is obtained from \pt{}-differential jet yields measured with respect to the experimentally accessible event plane \ep{}, which is reconstructed at forward rapidities ($2.8 < \eta < 5.1$ and  $-3.7 < \eta < -1.7$, Sec.~\ref{sec:ep}). The reported \vj{} has been corrected back to the azimuthal anisotropy with respect to the underlying symmetry plane \rptwo{} by applying an event plane resolution correction (Sec.~\ref{sec:233}). Jets are reconstructed at mid-rapidity ($\vert \eta_{\rm jet} \vert < 0.7$) using charged constituent tracks with momenta $0.15 < \pt{} < 100$ \GeVc{}, and are required to contain a charged hadron with \pt{} $\geq$ 3 \GeVc{}. The in-plane and out-of-plane jet spectra are unfolded independently to take into account detector effects and remaining azimuthally-dependent fluctuations in the underlying event transverse momentum density (Sec.~\ref{sec:23}). The jet spectra are corrected back to particle-level jets consisting of only primary charged particles from the collision. 

\section{Experimental setup and data analysis}
ALICE is a dedicated heavy-ion experiment at the LHC at CERN. A full overview of the detector layout and performance can be found in \cite{review,performance}. The central barrel detector system, covering full azimuth, is positioned in a solenoidal magnet with a field strength of 0.5 T. It comprises the Inner Tracking System (ITS) built from six layers of silicon detectors (the Silicon Pixel, Drift, and Strip Detectors: SPD, SDD and SSD) and a Time Projection Chamber (TPC). The two inner layers of the ITS, which comprise the SPD, are located at 3.9 and 7.2 cm radial distance from the beam axis.

The data presented in this paper were recorded in the Pb--Pb data taking periods in 2010 and 2011 at \sqrtSnn~= 2.76 TeV, using a minimium-bias trigger (2010) or an online centrality trigger for hadronic interactions (2011), which requires a minimum multiplicity in both the V0A and V0C detectors (discs of segmented scintillators covering full azimuth and $2.8 < \eta < 5.1$ and  $-3.7 < \eta < -1.7$, respectively). The V0 detectors are used to determine event centrality based on the energy deposition in the scintillator tiles \cite{centrality} and the event plane orientation, see Sec.~\ref{sec:ep}. Centrality, determined from the sum of the V0 amplitudes, is expressed as percentiles of the total hadronic cross section, with 0--5\% referring to the most central (largest multiplicity) events \cite{centrality}. The trigger is fully efficient in azimuth in the presented centrality ranges. Centrality estimation using the V0 system does not bias the \epn{} determination \cite{pidv}. Time information from the V0 detectors is used to reject beam-gas interactions from the event sample and the remaining contribution of such interactions is negligible. Only events with a primary vertex position within $\pm 10$~cm along the beam direction from the nominal interaction point were used in the analysis. A total of 6.8$\times$10$^6$ events with 0--5\%  centrality and 8.6$\times$10$^6$ events with 30--50\% centrality, corresponding to integrated luminosities of 18 and 5.6 $\mu$b$^{-1}$, respectively, are used in this analysis.

Charged particle tracks in this analysis are measured by the ITS and TPC and are selected in a pseudorapidity range $\vert\eta\vert<0.9$ with transverse momenta 0.15~$<$~\pt~$<$~100~\GeVc. To ensure a good momentum resolution, tracks were required to have at least three hits per track in the ITS. Since the SPD acceptance is non-uniform in azimuth for the data sample used in this analysis, two classes of tracks are used. The first class requires at least three hits per track in the ITS, with at least one hit per track in the SPD. The second class contains tracks without hits in the SPD, in which case the primary interaction vertex is used as an additional constraint for the momentum determination. For each track, the expected number of TPC space points is calculated based on its trajectory; tracks are accepted if they have at least 80\% of the expected TPC space-points, with a minimum of 70 TPC points. Tracks produced from interactions between particles and the detector, as well as tracks originating from weak decays (`secondary tracks') are rejected. The contribution of secondary tracks to the track sample is less than 10\% for tracks with \pt{} $<$ 1 \GeVc{} and negligible for tracks with higher transverse momentum.

\subsection{Event plane determination}\label{sec:ep}

The coefficient \vj{} quantifies azimuthal anisotropy with respect to \rptwo{}. The azimuthal anisotropy of the underlying event (`background flow') is also described by a Fourier series with harmonics $v_n = \langle \cos ( n [ \varphi - \rpn{}])\rangle$ \cite{Ollitrault:1992bk, zhang} where $\varphi$ denotes the track azimuthal angle. However, since the initial distribution of nucleons is not accessible experimentally, the \emph{event plane angles} \epn, \textit{i.e.} the axes of symmetry of the density of outgoing particles in the transverse plane, are used in place of \rpn{} when measuring \vj{} and $v_n$. 

The event plane angles \ep{} and \epthree{} in this study, corresponding to the two dominant Fourier harmonics, are reconstructed using the V0 detectors. Each V0 array consists of four rings in the radial direction, with each ring comprising eight cells with the same azimuthal size. The calibrated amplitude of the signal in each cell, proportional to the multiplicity incident on the cell, is used as a weight $w_{\mathrm{cell}}$ in the construction of the flow vectors $Q_n$ \cite{ep}

\begin{equation} \label{eq:qvect}
    Q_n = \sum_{\mathrm{cells}} w_{\mathrm{cell}} \exp \left( i~n~\varphi_{\mathrm{cell}} \right).
\end{equation}
In order to account for a non-uniform detector response which can generate a bias in the \epn{} azimuthal distribution, the components of the $Q_n$-vectors are adjusted using a re-centering procedure \cite{vzero, twist}. The V0A and V0C detectors cover different $\eta$ regions in which multiplicity $N$ and background flow $v_n$ may differ. The total V0 $Q$-vector is therefore constructed using weights $\chi_{n}$ \cite{ep} that are approximately proportional to the event plane resolution in each detector,
\begin{equation}
    Q_{n, {\mathrm{V0}}} = \chi^2_{n, \mathrm{V0A}} Q_{n, {\mathrm{V0A}}} + \chi^2_{n, \mathrm{V0C}} Q_{n, {\mathrm{V0C}}},
\end{equation}
to achieve the optimal combined event plane resolution.
The event planes are reconstructed from the real and imaginary parts of $Q_n$ as
\begin{equation}\label{eq:ep}
    \epn = \mbox{arctan} \left( \frac{\Im \left[Q_n\right]}{\Re \left[Q_n\right]}  \right) / n.
\end{equation}

The \vj{} itself is measured with respect to the second harmonic event plane angle. It is corrected for the finite precision with which the true symmetry plane is measured in the V0 system by applying an event plane resolution correction, see Sec.~\ref{sec:233}.

\subsection{Jet reconstruction in the presence of background flow}\label{sec:21}
Jet finding is performed using the FastJet \cite{fastjet,fastjetmanual} implementation of the infrared and collinear safe \kt{} and anti-\kt{} sequential recombination algorithms using the \pt{} recombination scheme and taking massless jet constituents. The resolution parameter $R = 0.2$ determines the characteristic maximum distance of constituent tracks to the jet axis in the $\eta$--$\varphi$ plane. 

In heavy-ion collisions, a large combinatorial background is present from particles that are not related to the hard scattering that produced a given jet. This background is subtracted from each jet on an event-by-event basis. The anti-\kt{} algorithm is used to find \emph{signal jets}. A fiducial cut of $\vert\eta_{\mathrm{jet}}\vert<$ 0.7 is applied on the signal jets to ensure that all jets are fully contained within the ITS and TPC acceptances and edge effects are avoided. The contribution of \emph{combinatorial} (or `fake') jets (clustered underlying event energy) to the measured jet spectrum is reduced by requiring that reconstructed jets contain at least one charged particle with \pt~$>$ 3 \GeVc{} and have an area of at least 0.56 $\pi R^2$. These selection criteria leave the hard part of the jet spectrum unaltered while significantly reducing the number of combinatorial jets which stabilizes the unfolding procedure \cite{charged_jets_raa,jets_raa,marta}.

The \kt-algorithm is used to estimate the average transverse momentum density of the underlying event, \rhoav{}, on an event-by-event basis. The quantity \rhoav{} is the median of the distribution of $\ptraw/A$ (the ratio of transverse momentum to jet area) of reconstructed $R = 0.2$ \kt-jets, excluding the leading two jets from the sample as proposed in \cite{salam} and implemented in earlier ALICE jet studies \cite{charged_jets_raa,jets_raa,marta}. The \kt{} jets are required to lie within $\vert \eta_{\rm jet} \vert < 0.7$ and have an area $A > 0.01$. The jet area $A$ is determined by embedding a fixed number of near zero-momentum \emph{ghost particles} per event prior to jet finding; the number of ghost particles in each reconstructed jet then gives a direct measure of the jet area. A ghost density of 200 particles per unit area is used, so that approximately 25 ghost particles are clustered into a jet with a radius of 0.2.

\begin{figure}
    \begin{center}
        \includegraphics[width=.7\textwidth]{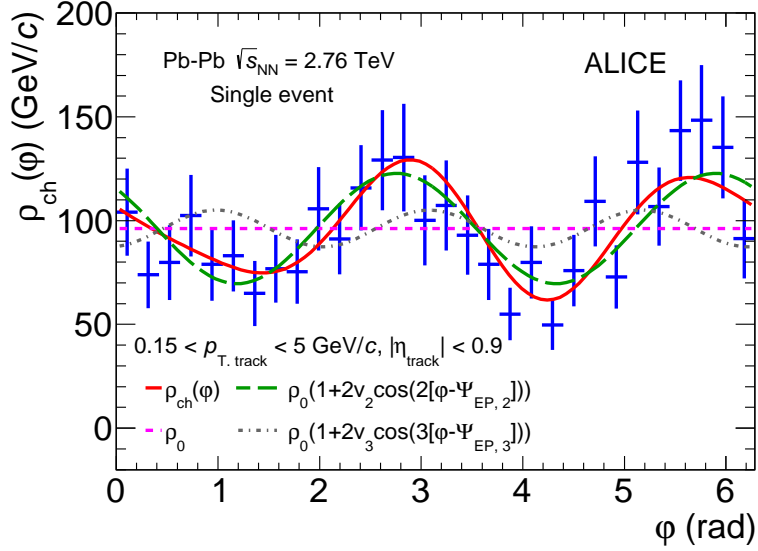}
        \caption{Transverse momentum density of charged tracks as a function of azimuthal angle for a single event from the most central 0-5\% event class. Data points (blue) are given with statistical uncertainties only. The red curve is the fit of Eq.~\ref{eq:master} to the distribution, the green and gray curves, obtained from the fit of Eq.~\ref{eq:master} as well, show the independent contributions of $v_2$ and $v_3$ to \rhovar{}. The dashed magenta line is the normalization constant $\rho_0$.}
    \label{fig:local_rho_fit}
    \end{center}
\end{figure}

In each event, the anisotropy of the underlying event is modeled using the dominant \cite{flowpaper} flow harmonics $v_2$ and $v_3$,
\begin{equation}\label{eq:master}
    \rhovar = \rho_0 \left(1+2 \lbrace v_2 \cos\left[2 \left( \varphi - \ep \right)\right] +v_3 \cos\left[3\left(\varphi - \epthree \right)\right] \rbrace \right).
\end{equation}
Here, \rhovar{} is the azimuthal distribution of summed track \pt{} for tracks with 0.15 $<$ \pt~ $<$ 5 \GeVc{} and $\vert \eta_{\mathrm{track}} \vert < $ 0.9. The parameters $\rho_0$ and $v_n$ are determined event-by-event from a fit of the right side of Eq.~\ref{eq:master} to the data. The event plane angles \epn{} are not fitted, but fixed to the V0 event plane angles. A single event example of this procedure is illustrated in Fig. \ref{fig:local_rho_fit}, where the data points represent the transverse momentum density distribution in a single event, the red curve represents the full functional description of \rhovar{} (Eq.~\ref{eq:master}), the green and gray curves give the contributions of the separate harmonics $v_2$ and $v_3$, and the dashed magenta line is the normalization constant $\rho_0$.  To reduce the bias of hard jets in the estimates of $v_n$ in Eq.~\ref{eq:master} while retaining azimuthal uniformity, the leading jet in each event is removed by rejecting all tracks for which $\vert \eta_{\rm jet} - \eta_{\rm track} \vert < R$. 
The $\eta$ separation between the tracks and the V0 detectors also removes short range correlations between the event planes and tracks. 

The number of bins to which Eq.~\ref{eq:master} is fitted is set on an event-by-event basis to the square root of the number of tracks. The fit maximizes the estimated likelihood \cite{baker}, which is based on a Poisson distribution for the bin content. Since the bin contents are not pure counts, but weighted by \pt, the statistical uncertainties on each bin $\sigma_i$ are estimated as the sum of the squares of the \pt{} of the individual particles: $\sigma_i = \sigma(\sum \pt) = \sqrt{} \sum \pt^2$. A scaled Poisson distribution $w_i  P(x_i\vert \mu_i/w_i)$ is used as the probability distribution for the data points in the likelihood calculation, with a scale factor $w_i = \sigma_i^2 / y_i$ where $y_i$ is the bin content and $\mu_i/w_i$ is the expected signal from the fit function. The compatibility of each fit with the data is tested by calculating the $\chi^2$ and evaluating the probability of finding a test statistic at least as large as the observed one in the $\chi^2$ distribution. When this probability is less than 0.01, the average event background density \rhoav~is used instead of \rhovar; 
this occurs in 3\% (most central) to 7\% (semi-central)  of events. The acceptance criterion is varied in the systematic studies; the sensitivity to it is small.

\begin{figure}
    \includegraphics[width=.5\textwidth]{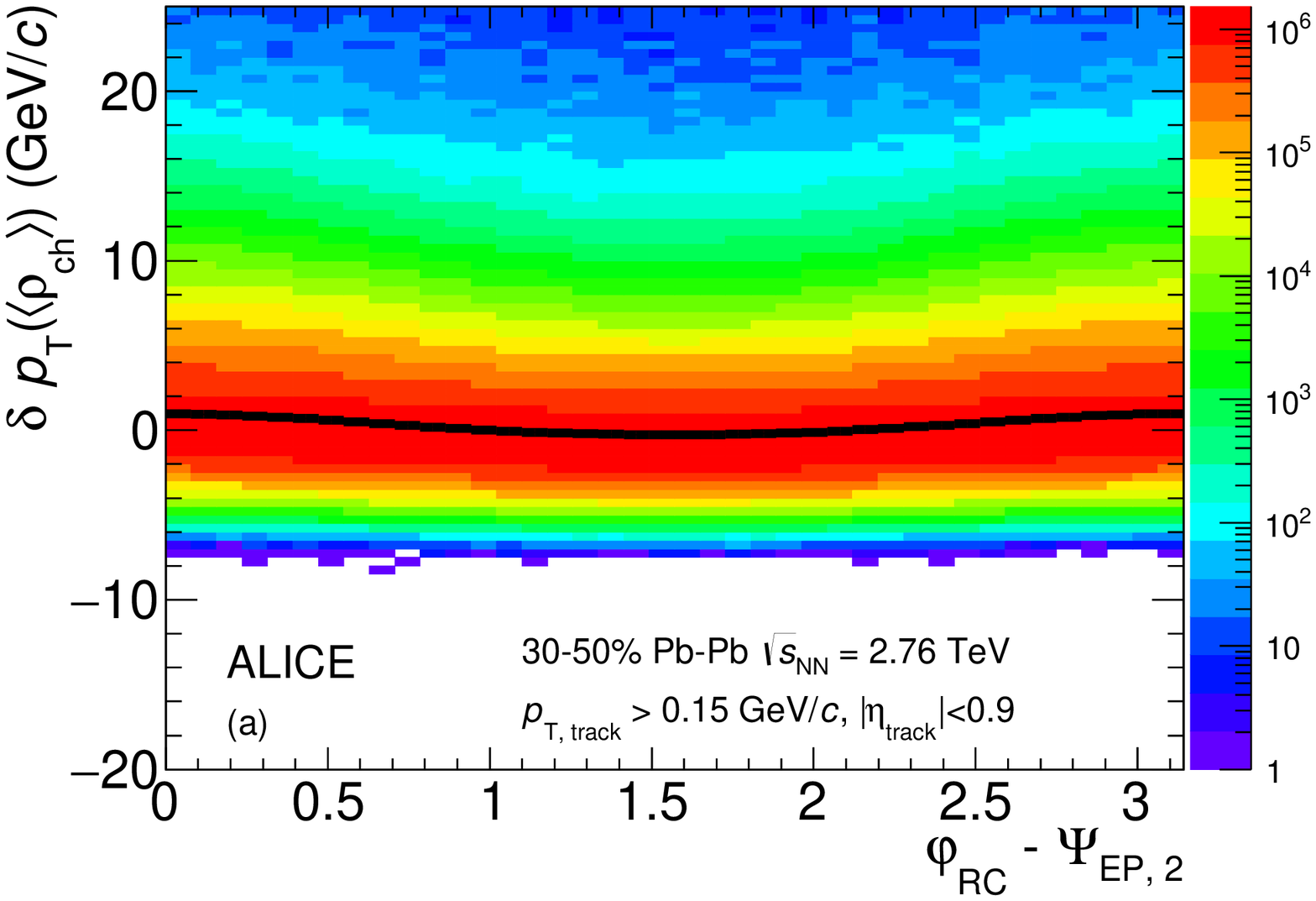}
    \includegraphics[width=.5\textwidth]{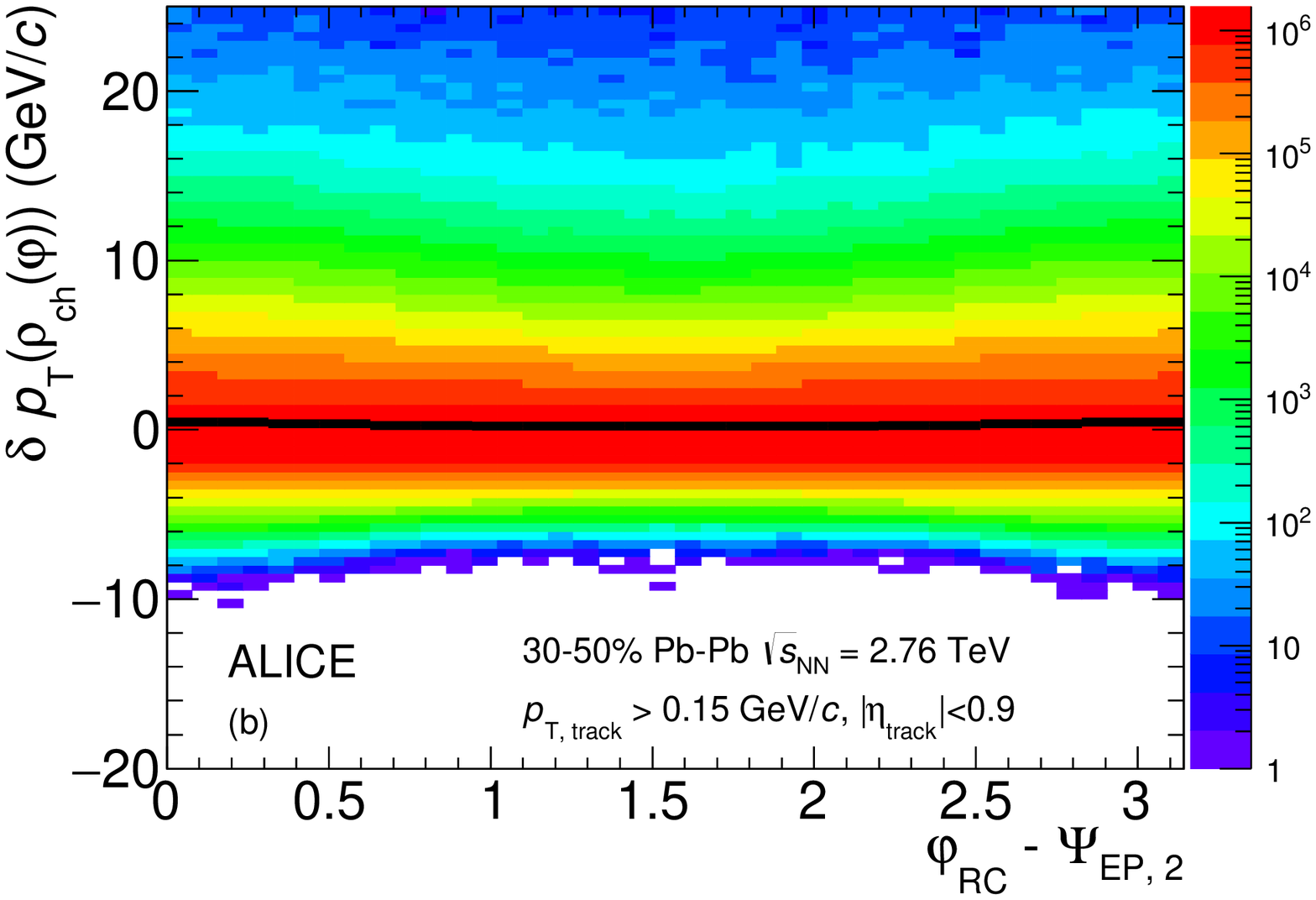}
    \caption{The \dpt~distribution (Eq. \ref{eq:dpt}) from the random cone (RC) procedure as function of cone azimuthal angle $\varphi_{\mathrm{RC}}$ relative to the event plane. In panel (a) the azimuthally-averaged background \rhoav{} has been subtracted; in panel (b) the azimuthally dependent \rhovar{} from an event-by-event fit of the $\pt$-density with Eq. \ref{eq:master}. The solid black line represents the mean of the \dpt{} distribution. 
}
    \label{fig:tech_2}
\end{figure}

The corrected transverse momentum \ptjet~of a jet of area $A$ is calculated from the measured raw jet momentum, \ptraw, as 
\begin{equation}\label{master_jet}
    \ptjet = \ptraw - \rholocal \, A 
\end{equation}
where  \rholocal~is obtained from integration of \rhovar{} around $\phijet \pm R$ 
\begin{equation}\label{eq:int}
    \rholocal = \frac{\rhoav}{2 R \rho_0} \int_{\varphi - R}^{\varphi + R} \rhovar d \varphi.
\end{equation}
The pre-factor of the integral, $\frac{\rhoav}{2 R \rho_0}$, is chosen such that integration over the full azimuth yields the average transverse momentum density $\rhoav$.
The validity of Eq.~\ref{eq:master} as a description of the contribution of background flow to the underlying event energy is tested by placing cones of radius $R=0.2$ at random positions (excluding the location of the leading jet) in the $\eta$--$\phi$ plane and subtracting the expected summed transverse momentum in a cone from the measured transverse momentum in the cone,
\begin{equation}\label{eq:dpt}
    \dpt = \sum \pt^{\mathrm{tracks}} - \rho \pi R^2.
\end{equation}
Here, $\rho$ is the expected transverse momentum density. This procedure is repeated multiple times per event, until the full phase space is covered, to obtain a  distribution of \dpt~values. The \dpt~distribution as a function of the cone azimuthal angle $\varphi_{\mathrm{RC}}$ relative to the event plane \ep{} is shown in Fig. \ref{fig:tech_2}. In panel (a) \rhoav{} has been used for the estimation of the underlying event summed \pt{} and in panel (b) \rhovar. Incorporating azimuthal dependence into the underlying event description leads to a sizable reduction in the cosine modulation of the \dpt~distribution. 

The effectiveness of the subtraction of background flow is quantified by comparing the expected and measured widths of the \dpt~distribution in the \emph{absence} of background flow, $\sigma$(\dpt$^{v_n=0}$), (see Fig. \ref{fig:tech_2}b) to the expected and measured widths of the \dpt~distribution in the \emph{presence}  of background flow, $\sigma$(\dpt$^{v_n}$) (Fig. \ref{fig:tech_2}a). Assuming independent particle emission and Poissonian statistics, the expected width of the \dpt~distribution in the absence of background flow ($v_n=0$) is given by \cite{marta}
\begin{equation}\label{eq:poisson_only}
    \sigma (\dpt^{v_n=0}) = \sqrt{N_A  \sigma^2(\pt) + N_A  \langle \pt \rangle^2 }
\end{equation}
where $N_A$ is the average expected number of tracks within a cone, $\langle \pt \rangle$ the mean \pt~ of a single particle spectrum and $\sigma(\pt)$ the standard deviation of this spectrum. This expectation can be extended to include contributions from background flow by introducing non-Poissonian density fluctuations (the background flow harmonics $v_n$) \cite{marta}, as
\begin{equation}\label{eq:poisson_and_more}
    \sigma (\dpt^{v_n}) = \sqrt{N_A  \sigma^2(\pt) + (N_A + 2 N_A^2 (v_2^2 + v_3^2
)) \langle \pt \rangle^2 }.
\end{equation}

\begin{figure}
    \begin{center}         
        \includegraphics[width=.7\textwidth]{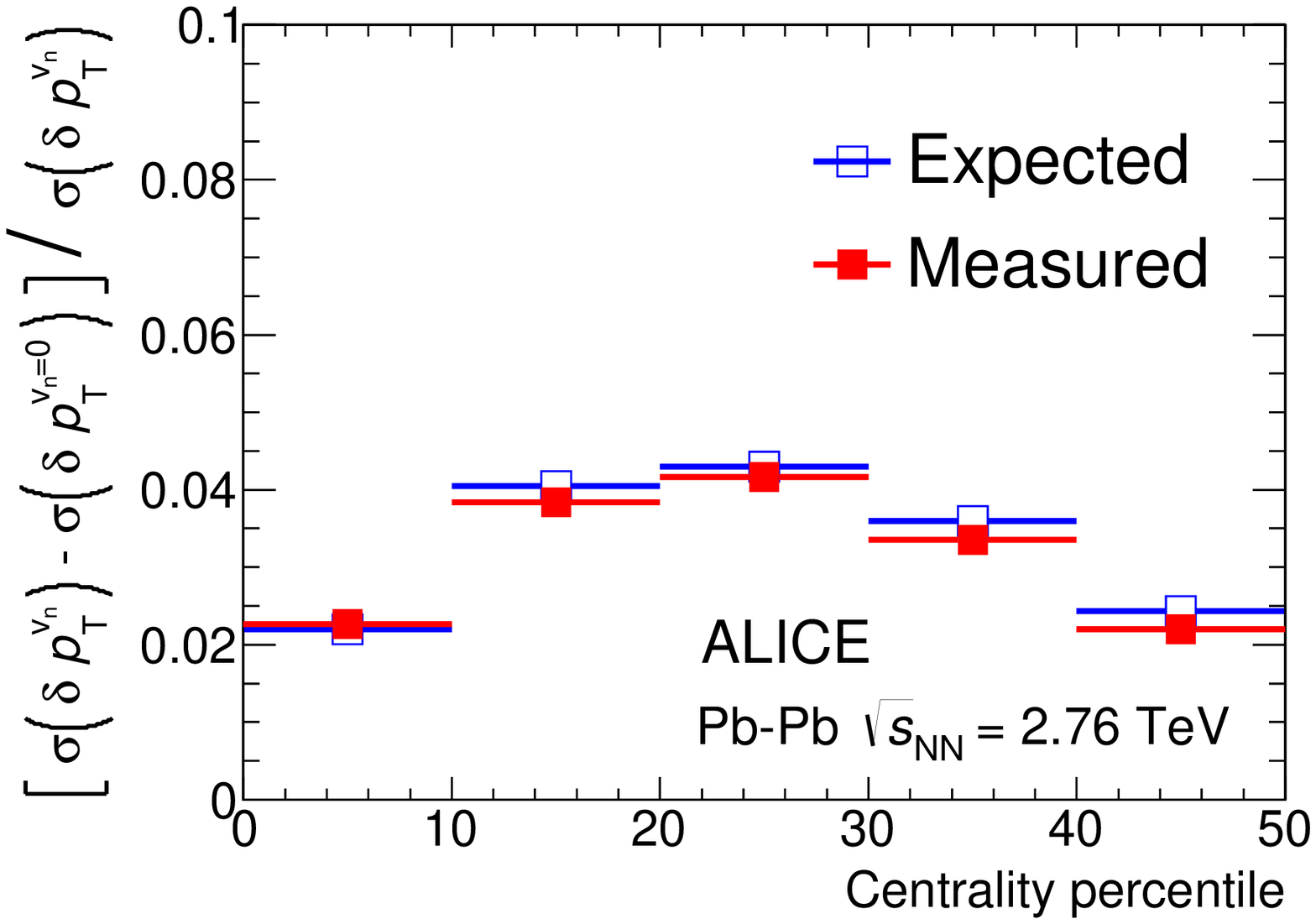}
        \caption{Centrality dependence of the measured and expected relative change in the \dpt{} distribution width from using the azimuthally dependent \rholocal{} instead of the median \rhoav{}. The blue points give the expected reduction from simple assumptions about the behavior of charged particle spectra and flow harmonics $v_n$ (following Eq.~\ref{eq:poisson_only} and \ref{eq:poisson_and_more}). The red points use the measured widths from \dpt{} distributions directly. Statistical uncertainties are smaller than the marker size.
        }

        \label{fig:improvements}
    \end{center}
\end{figure}
The measured widths are obtained from the \dpt{} distributions directly; the distributions are constructed using as the transverse momentum density $\rho$ in Eq.~\ref{eq:dpt} either \rhoav{} to obtain $\sigma(\dpt^{v_n})$ or \rholocal{} for $\sigma(\dpt^{v_n=0})$. 

Figure \ref{fig:improvements} shows the expected and measured relative change in the width of the \dpt~distribution, quantified as $(\sigma(\dpt^{v_n}) - \sigma(\dpt^{v_n=0})) / \sigma(\dpt^{v_n})$, as function of collision centrality. The blue points give the expected reduction from 
Eqs~\ref{eq:poisson_only} and \ref{eq:poisson_and_more}. The red points use the measured widths from \dpt{} distributions. The expected change is in good quantitative agreement with the measured change over the entire centrality range, indicating that the width of the \dpt{} distributions can be understood in terms of a simple independent particle emission model with background flow contributions.

The background subtraction, unfolding, correction for the reaction plane resolution as described in Sections \ref{sec:23} and \ref{sec:233} were also validated using events consisting of PYTHIA jets embedded in heavy-ion background events and toy model events. In the first study, full PYTHIA pp events were combined with reconstructed Pb--Pb collisions to create events with a controlled signal and background. The signal jets from PYTHIA have no preferred orientation, $\vj=0$, while the heavy-ion events have a non-zero $v_{2}$ of the soft particles. Jets found in the events were matched to the embedded PYTHIA jets and the analysis was carried out with matched jets only. After unfolding, the $\vj$ was compatible with 0, as expected. The other study was based on events generated using a simple thermal model for soft particle production and a distribution of high-\pt{} particles that resembles the jet spectrum, as suggested in \cite{lh}. A non-zero $v_{2}=0.07$ was introduced for momenta $\pt < 5$ \GeVc{} to model the background flow and two variations at large $\pt > 30$ \GeVc : $v_{2}=0$ or $v_{2}=0.05$. In both cases, the input flow values were correctly reconstructed by the analysis.

\subsection{Unfolding}\label{sec:23}
After the subtraction procedure presented in the previous section, the measured jet spectrum is unfolded \cite{blobel,schmelling} to correct for detector effects and fluctuations in the underlying event transverse momentum density. Mathematically, the unfolded jet spectrum can be derived from the measured spectrum by solving 
\begin{equation}\label{unfolding_analysical}
    M ( \ptrec ) = \int  G ( \ptrec, \ptgen ) T ( \ptgen ) \epsilon ( \ptgen ) \mbox{d} \ptgen
\end{equation}
for $T ( \ptgen )$, the unfolded true jet spectrum, where $M ( \ptrec )$ is the measured jet spectrum, $G$(\ptrec,\ptgen) is a functional description (\emph{response function}) of distortions due to background fluctuations and detector response,  and $\epsilon ( \ptgen )$ is the jet finding efficiency. The coefficient \vj{} is not affected by the efficiency, hence $\epsilon ( \ptgen )$  will be omitted from here on. Since the measured jet spectrum is binned, Eq.~\ref{unfolding_analysical} is discretized by replacing the integral by a matrix multiplication
\begin{equation}\label{unfolding_matrix}
    \mbox{\textbf{M}}_{\mbox{\textbf{m}}} = \mbox{\textbf{G}}_{\mbox{\textbf{m}},\mbox{\textbf{t}}}\cdotp \mbox{\textbf{T}}_{\mbox{\textbf{t}}}^{\prime}
\end{equation}
where $\mbox{\textbf{T}}_{\mbox{\textbf{t}}}^{\prime}$ is the solution of the discretized equation (the prime indicates that $\mbox{\textbf{T}}_{\mbox{\textbf{t}}}^{\prime}$ is not corrected for jet-finding efficiency). The combined response matrix $\mbox{\textbf{G}}_{\mbox{\textbf{m}},\mbox{\textbf{t}}}$ is the product of the response matrices from detector effects and transverse momentum density fluctuations, the latter of which are constructed independently for the in-plane and out-of-plane spectra by embedding random cones at specific relative azimuth with respect to the event plane (see the text below Eq.~\ref{eq:jetflow} for the definition of the intervals). 

The detector response matrix is obtained by matching pp jets generated by PYTHIA \cite{pythia} (`particle-level' jets) to the \emph{same} jets after transport through the detector (`detector-level' jets) by GEANT3 \cite{geant}, where the detector conditions are tuned to those of the Pb--Pb data-taking periods. Particle-level jets contain only primary charged particles produced by the event generator, which comprise all prompt charged particles produced in the collision, as well as products of strong and electromagnetic decays, while products of weak decays of strange hadrons are rejected. Matching is based on the shortest distance in the $\eta$--$\varphi$ plane between detector level and particle level jets and is bijective, meaning that there is a one-to-one correspondence between detector and particle level jets. The response matrix for background fluctuations is constructed from the \dpt~distributions, which, when normalized, are probability distributions for the change of the jet energy caused by background fluctuations.

Solving Eq.~\ref{unfolding_matrix} requires inversion of $\mbox{\textbf{G}}_{\mbox{\textbf{m}}, \mbox{\textbf{t}}}$ and generally leads to non-physical results which oscillate wildly due to the statistical fluctuations of the measured jet yield. The unfolded solution therefore needs to be regularized. In general this is done by introducing a penalty term for large local curvatures associated with oscillations. Various algorithms for regularized unfolding exist; the unfolding method based on the Singular Value Decomposition (SVD unfolding) \cite{svdunf} is used in this study. A comparison to the unfolded solution from $\chi^2$ minimization \cite{chi2} is used in the systematic studies. 

The measured jet spectrum is taken as input for the unfolding routine in the range $30 < \ptjet < 105$ \GeVc~ for 0--5\% collision centrality and 15 $< \ptjet < 90$ \GeVc~ for 30--50\% collision centrality. The lower bound corresponds to five times the width of the \dpt~distribution, the upper bound is the edge of the last measured bin which contains at least 10 counts. This configuration was found to lead to reliable unfolded solutions in Monte Carlo studies \cite{lh,charged_jets_raa}. The unfolded jet spectrum starts at 0 \GeVc~ to allow for feed-in of true jets with low \ptjet{}. In addition, combinatorial jets which are not rejected by the jet area and leading charged particle requirements are migrated to momenta lower than the minimum measured \ptjet. The unfolded solution ranges up to 200 \GeVc~(0--5\%) and 170~\GeVc~(30--50\%) to allow for migration of jets to a \ptjet{} higher than the maximum measured momentum. As the data points of the unfolded solution are strongly correlated for \ptjet{} outside the experimentally measured interval, \vj{} will be reported only within the limits of the measured jet spectra.

\subsection{Evaluation of \vj}\label{sec:233}
The coefficient \vj{} is calculated from the difference between the unfolded \pt-differential jet yields in-plane (\nin) and out-of-plane (\nout) with respect to the second harmonic event plane, corrected for event plane resolution,
\begin{equation}\label{eq:jetflow}
    \vj(\ptjet) = \frac{\pi}{4}  \frac{1}{\mathcal{R}_2} \frac{\nin(\ptjet) - \nout(\ptjet) }{\nin(\ptjet) +\nout(\ptjet) }.
\end{equation}
Eq.~\ref{eq:jetflow} is derived by integrating Eq.~\ref{eq:flowdud} for $n = 2$, over intervals $\left[ -\frac{\pi}{4}, \frac{\pi}{4} \right]$ and  $ \left[ \frac{3 \pi}{4}, \frac{5\pi}{4} \right]$ for \nin~and $\left[ \frac{\pi}{4}, \frac{3\pi}{4}\right]$ and $\left[\frac{5 \pi}{4}, \frac{7\pi}{4}\right]$ for \nout, substituting \ep{} for \rptwo{}. Eq.~\ref{eq:jetflow} is sensitive to correlations between even-order harmonics $v_{2n}$ and \ep{}. As a result of the integration limits however, the first harmonic of the Fourier expansion that can contribute to the observed \vj{} is \vjsix{}. The V0 event plane resolution $\mathcal{R}_2$ is introduced to account for the finite precision with which the true symmetry plane \rptwo{} is measured in the V0 system and is defined as 
\begin{equation}
    \mathcal{R}_2 = \left< \cos \left[ 2 \left( \ep^{\mathrm{V0}} - \rptwo \right) \right] \right>.
\end{equation}
Measuring event planes in multiple $\eta$ regions (\emph{sub-events}) allows for the evaluation of the resolution directly from data \cite{3sub,3suberr}. Using the full V0 detector and negative and positive $\eta$ sides of the TPC as sub-events, the resolution in Eq.~\ref{eq:jetflow} is evaluated as
\begin{equation}
    \mathcal{R}_2 = \left( \frac{\left< \cos \left[ 2 \left( \ep^{\mbox{\tiny{V0}}} - \ep^{\mbox{\tiny{TPC, $\eta >$ 0}}} \right) \right] \right>   \left< \cos \left[ 2 \left( \ep^{\mbox{\tiny{V0}}} - \ep^{\mbox{\tiny{TPC, $\eta <$ 0}}} \right) \right] \right>  }{ \left< \cos \left[ 2 \left( \ep^{\mbox{\tiny{TPC, $\eta >$ 0}}} - \ep^{\mbox{\tiny{TPC, $\eta <$ 0}}} \right) \right] \right>  } \right)^{1/2}.
\end{equation}
The event plane resolution $\mathcal{R}_2$ is found to be 0.47 in 0--5\% centrality and 0.75 in 30--50\% centrality with negligible uncertainties. The \ep{} angles in the TPC are obtained following the procedure of Eq.~\ref{eq:ep} on tracks with $0.15 < \pt{} < 4\ \GeVc{}$, using unit track weights in the construction of the flow vectors $Q_2$ (see Eq. \ref{eq:qvect}).

Using the V0 detectors for the reconstruction of the event plane guarantees that the jet axis and event plane information are separated in pseudorapidity by $\vert \Delta \eta \vert >$ 1 and thus removes autocorrelation biases between the signal jets and event plane orientation. 
The possible non-flow correlation between the event plane angle and jets due to  di-jets with one jet at mid-rapidity and one jet in the V0 acceptance was studied using the PYTHIA event generator. The rate of such di-jet configurations was found to be negligible (less than 1 per mille of the total di-jet rate at mid-rapidity) for $\ptjet>20$ GeV. Possible effects from back-to-back jet pairs with a jet in each of the V0 detectors are even smaller.

\subsection{Systematic uncertainties}\label{sec:24}

The measured \vj{} is corrected for experimental effects, such as the finite event plane resolution and detector effects on the jet energy scale as well as the effects of the uncorrelated background and its fluctuations using the corrections outlined in the Sections \ref{sec:ep}--\ref{sec:233}. Hydrodynamic flow of the background is taken into account event-by-event in the underlying event description, residual effects are removed by azimuthally independent unfolding. The remaining uncertainties in these correction procedures are treated as systematic uncertainties. Systematic uncertainties on \vj~are grouped into two categories, \emph{shape} and \emph{correlated}, based on their point-to-point correlation. Shape uncertainties are anti-correlated between parts of the unfolded spectrum: when the yield in part of the spectrum increases, it decreases elsewhere and vice versa. Correlated uncertainties are correlated point-to-point. Both types of uncertainties however have contributions which lead to correlated changes of \nin~and \nout.

Correlated uncertainties are estimated for the in-plane and out-of-plane jet spectra independently. Two sources of correlated uncertainties are considered: tracking efficiency and the inclusion of combinatorial jets in the measured jet spectrum. The dominant correlated uncertainty ($ \lesssim$ 10\%) arises from tracking and is estimated by constructing a detector response matrix with a tracking efficiency reduced by 4\% (motivated by studies \cite{charged_jets_raa} comparing reconstructed tracks to simulations of HIJING \cite{hijing} events). The observed difference between the nominal and modified unfolded solution is taken as a symmetric uncertainty to allow for an over- and underestimation of the tracking efficiency. The sensitivity of the unfolded result to combinatorial jets is tested by changing the lower range of the unfolded solution from 0 to 5 \GeVc, which leads to an overall (correlated) increase of the unfolded jet yield. Both correlated uncertainties are added in quadrature and propagated to \vj~assuming that variations are strongly correlated between the in-plane and out-of-plane jet spectra, while still allowing for effects from azimuthally-dependent variations in track occupancy and reconstruction efficiency, by setting the sample correlation coefficient $\rho \equiv \sigma_{i, j}/(\sigma_i \sigma_j)$ to 0.75. 

Shape uncertainties fall into three categories: assumptions in the unfolding procedure, feed-in of combinatorial jets, and the sensitivity of the unfolded solution to the shape of the underlying event energy distribution. The dominant contribution to the unfolding uncertainty is related to the regularization of the unfolded solution. The SVD algorithm \cite{svdunf} regularizes the unfolding by omitting components of the measured spectrum for which the singular value is small and which amplify statistical noise in the result. To explore the sensitivity of the result to the regularization strength, the effective rank of the matrix equation that is solved is varied by changing an integer regularization parameter $k$ by $\pm$ 1. The SVD unfolding algorithm uses a prior spectrum as the starting point of the unfolding; the result of the unfolding is the ratio between the full spectrum and this prior. The unfolded solution from the $\chi^2$ algorithm \cite{chi2} is used as prior (default) as well as a PYTHIA spectrum. The bias from the choice of unfolding algorithm itself is tested by comparing the results of the SVD unfolding and the $\chi^2$ algorithm. 

The same nominal unfolding approach is used for the in-plane and out-of-plane jet spectra and the \dpt~distributions for the in-plane and out-of-plane background fluctuations are similar in width; the unfolding uncertainty is therefore strongly correlated between the in-plane and out-of-plane jet spectra. These correlations are taken into account by applying the variations in the unfolding procedure to the in-plane and out-of-plane jet spectra at the same time and calculating the resulting variations of \vj{}. The total uncertainty from unfolding is determined by constructing a distribution of all unfolded solutions in each \ptjet{} interval and assigning the width of this distribution as a systematic uncertainty. 

The other two components of the shape uncertainty are the sensitivity of the unfolded solution to combinatorial jets and uncertainties arising from the description of the underlying event; both are estimated on the in-plane and out-of-plane jet spectra independently and propagated to \vj~as uncorrelated. A systematic uncertainty is only assigned when the observed variation is found to be statistically incompatible with the nominal measurement. The effect of combinatorial jets is tested by varying the minimum \ptjet{} of the measured jet spectrum by $\pm$ 5 \GeVc, effectively increasing or decreasing the possible contribution of combinatorial jet yield at low jet momentum. 
To test the assumptions made in the fitting of Eq.~\ref{eq:master} the maximum \pt{} of accepted tracks is lowered to 4 \GeVc. Additionally, the minimum $p$-value that is used as a goodness of fit criterion is changed from 0.01 (the nominal value) to 0.1. The minimum required distance of tracks to the leading jet axis in pseudorapidity is enlarged to 0.3. 

Table~\ref{tab:syst05} gives an overview of the systematic uncertainties in terms of absolute uncertainties on \vj~for all sources (where the total uncertainty is the quadratic sum of the separate components). High statistics Monte Carlo testing has been used to verify that uncertainties labeled `$\ll$ stat' are indeed negligible compared to other uncertainties.

%
%

\begin{table}
\begin{center}
\begin{tabular}{r||l||r|r|r||r|r|r}
\hline 
 & &  \multicolumn{6}{c}{Uncertainty on \vj} \\
\hline
&  \ptjet~(GeV/$c$)  & 30--40&60--70&80--90 &30--40&60--70  &80--90 \\
\hline
\hline
& Centrality (\%)& \multicolumn{3}{c}{0--5}  & \multicolumn{3}{c}{30--50 }  \\

\hline
\multirow{3}{*}{Shape $ $ } & Unfolding & 0.017& 0.012&  0.016 & 0.016& 0.011&0.015 \\
                            & \ptjet-measured & 0.013& $\ll$ stat & $\ll$ stat & 0.024& $\ll$ stat& $\ll$ stat\\
                            & \rhovar fit & 0.015& $\ll$ stat& 0.016  &$\ll$ stat& $\ll$ stat& $\ll$ stat\\
\hline
Total & &0.027 & 0.012& 0.023 &0.029 &0.011 &0.015 \\
\hline 
\multirow{2}{*}{Correlated $ $ }& Tracking &0.009&0.009  &0.009  & 0.007& 0.007& 0.007\\
                                & \ptjet-unfolded & $\ll$ stat& $\ll$ stat&$\ll$ stat  & $\ll$ stat& $\ll$ stat& $\ll$ stat\\

\hline
Total & & 0.009& 0.009&  0.009& 0.007& 0.007& 0.007\\
\hline
\end{tabular}
\end{center}
\caption{Systematic uncertainties on \vj{} for various transverse momenta and centralities. Uncertainties in central and semi-central collisions are given in the same \pt{} ranges. The definitions of shape uncertainty and correlated uncertainty are explained in Sec.~\ref{sec:24}. Fields with the value `$\ll$ stat' indicate that no systematic effect can be resolved within the statistical limits of the analysis. }
\label{tab:syst05}
\end{table}

\section{Results and discussion}\label{sec:3}
\begin{figure}
    \includegraphics[width=.5\textwidth]{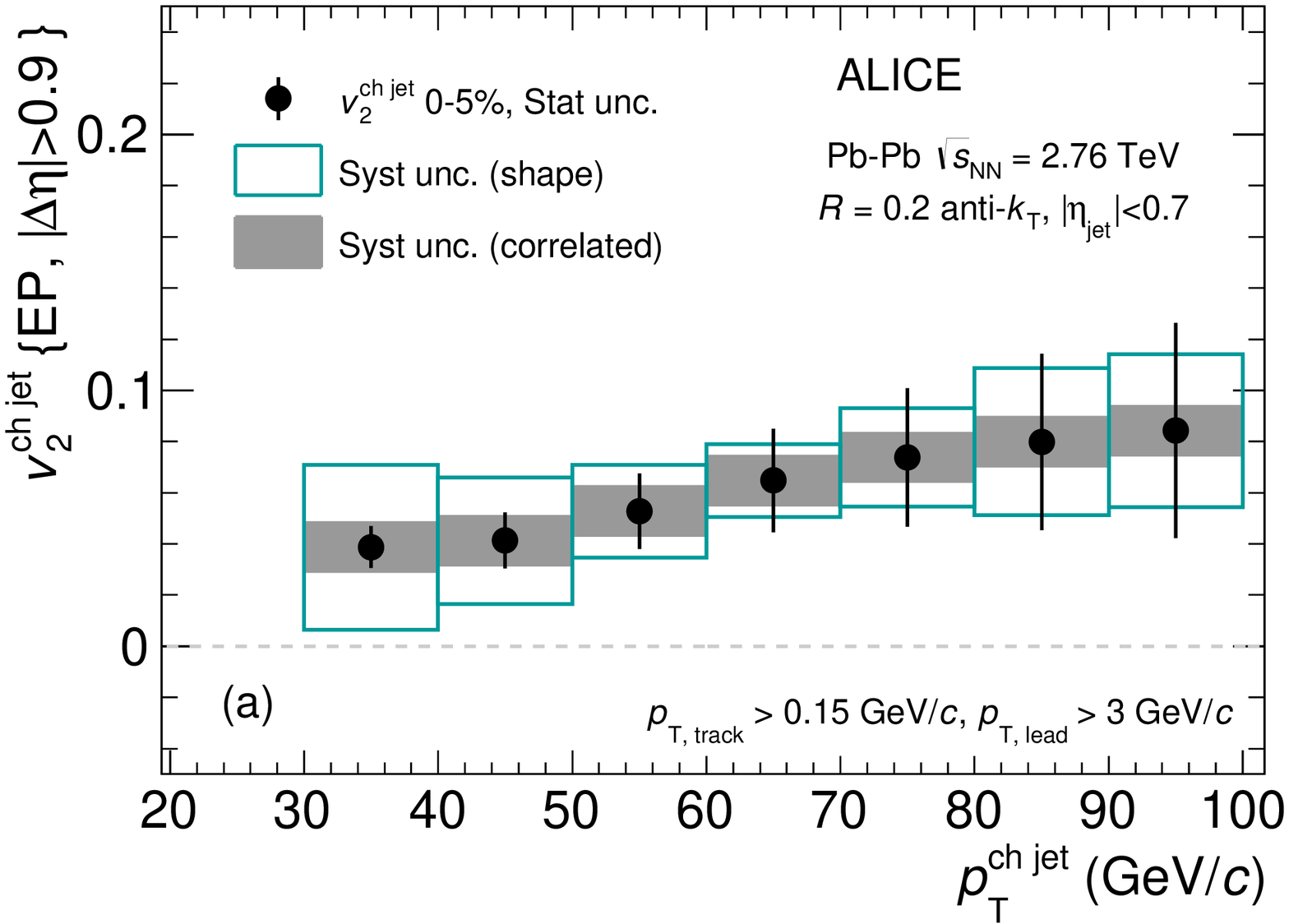}
    \includegraphics[width=.5\textwidth]{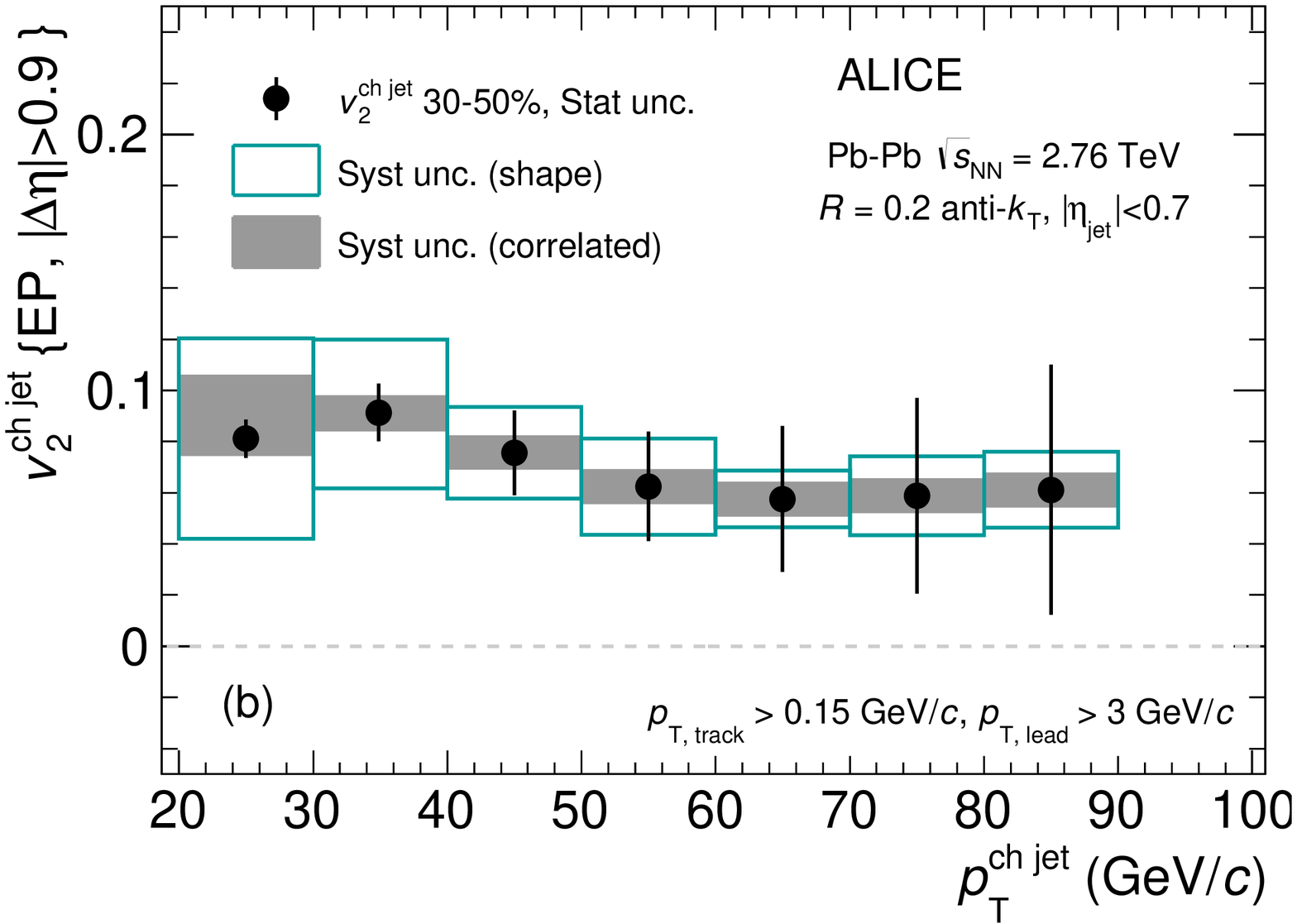}
    \caption{Second-order harmonic coefficient \vj~as function a of \ptjet{} for 0--5\% (a) and 30--50\% (b) collision centrality. The error bars on the points represent statistical uncertainties, the open and shaded boxes indicate the shape and correlated uncertainties (as explained in Sec.~\ref{sec:24}). }
    \label{fig:results}
\end{figure}

The coefficients \vj~as function of \ptjet{} for 0--5\% and 30--50\% collision centrality are presented in Fig.~\ref{fig:results}. Significant positive \vj~is observed in semi-central collisions and no (significant) \pt~dependence is visible. The observed behavior is indicative of path-length-dependent in-medium parton energy loss. The observed \vj{} in central collisions is of similar magnitude. The systematic uncertainties on the measurement however are larger than those on the semi-central \vj{} data, in particular at lower \ptjet, as a result of the larger relative background contribution to the measured jet energy. 

The significance of the results is assessed by calculating a $p$-value for the hypothesis that $\vj=0$ over the presented momentum range. The $p$-value is evaluated starting from a modified  $\chi^2$ calculation that takes into account both statistical and (correlated) systematic uncertainties, as suggested in \cite{phenix}. The modified  $\chi^2$ for the hypothesis $\vj=\mu_i$ is calculated by minimizing
\begin{equation}\label{a}
    \tilde{\chi}^2 (\epsilon_{\rm corr}, \epsilon_{\rm shape}) = \left[ \left( \sum_{i=1}^{n} \frac{(v_{2, i} + \epsilon_{\rm corr} \sigma_{\rm corr, i} + \epsilon_{\rm shape} - \mu_i)^2}{\sigma^2_i} \right)+ \epsilon_{\rm corr}^2 +  \frac{1}{n}\sum_{i=1}^{n} \frac{\epsilon_{\rm shape}^2}{\sigma^2_{\rm shape, i}}   \right]
\end{equation}
with respect to the systematic shifts $\epsilon_{\rm shape}$, $\epsilon_{\rm corr}$, where $v_{2, i}$ represent the measured data ($n$ points), $\sigma_i$ are statistical uncertainties and $\sigma_{\rm shape, i}$, $\sigma_{\rm corr, i}$ denote the two specific types of systematic uncertainties. The parameter $\epsilon_{\rm shape}$ is a measure of the fully correlated shifts; a shift of all data points by the correlated incertainty  $\sigma_{\rm corr, i}$  gives a total contribution to $\tilde{\chi}^2$ of one unit. The systematic shifts for the shape uncertainty are taken to be of equal size for each point, since this gives the best agreement with the $\vj=0$ hypothesis and thus provides a conservative estimate of the significance; the penalty factor is constructed such that an average shift of all data points by $\sigma_{\rm shape}$ adds one unit to  $\tilde{\chi}^2$.

The $p$-value itself is calculated using the $\chi^2$ distribution with $n-2$ degrees of freedom. For semi-central collisions a $p$-value of 0.0009 is found, indicating significant positive \vj{}. It should be noted that the most significant data points are at $\ptjet < 60$ \GeVc; the results in the range $60<\ptjet<100$ \GeVc{} are compatible with $\vj=0$ ($p$-value 0.02). 
In central collisions, a $p$-value with respect to the hypothesis of \vj{} = 0 of 0.12 is found which indicates that \vj~is compatible with 0 within two standard deviations. Following the same approach an upper limit of \vj{} = 0.088 is found within the same confidence interval. 

\subsection{Comparison to previous measurements and model predictions}\label{sec:32}
\begin{figure}
    \includegraphics[width=.5\textwidth]{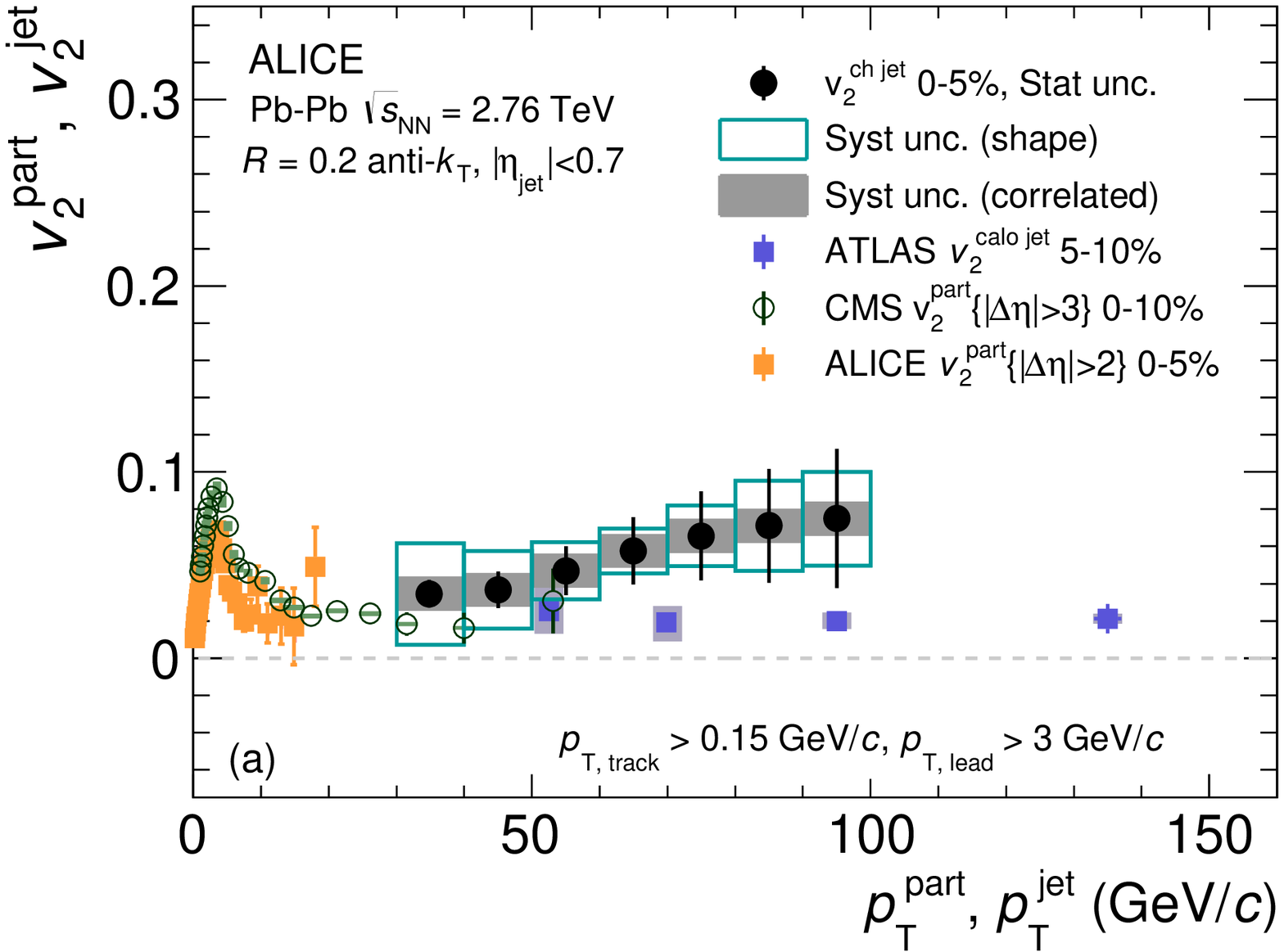}
    \includegraphics[width=.5\textwidth]{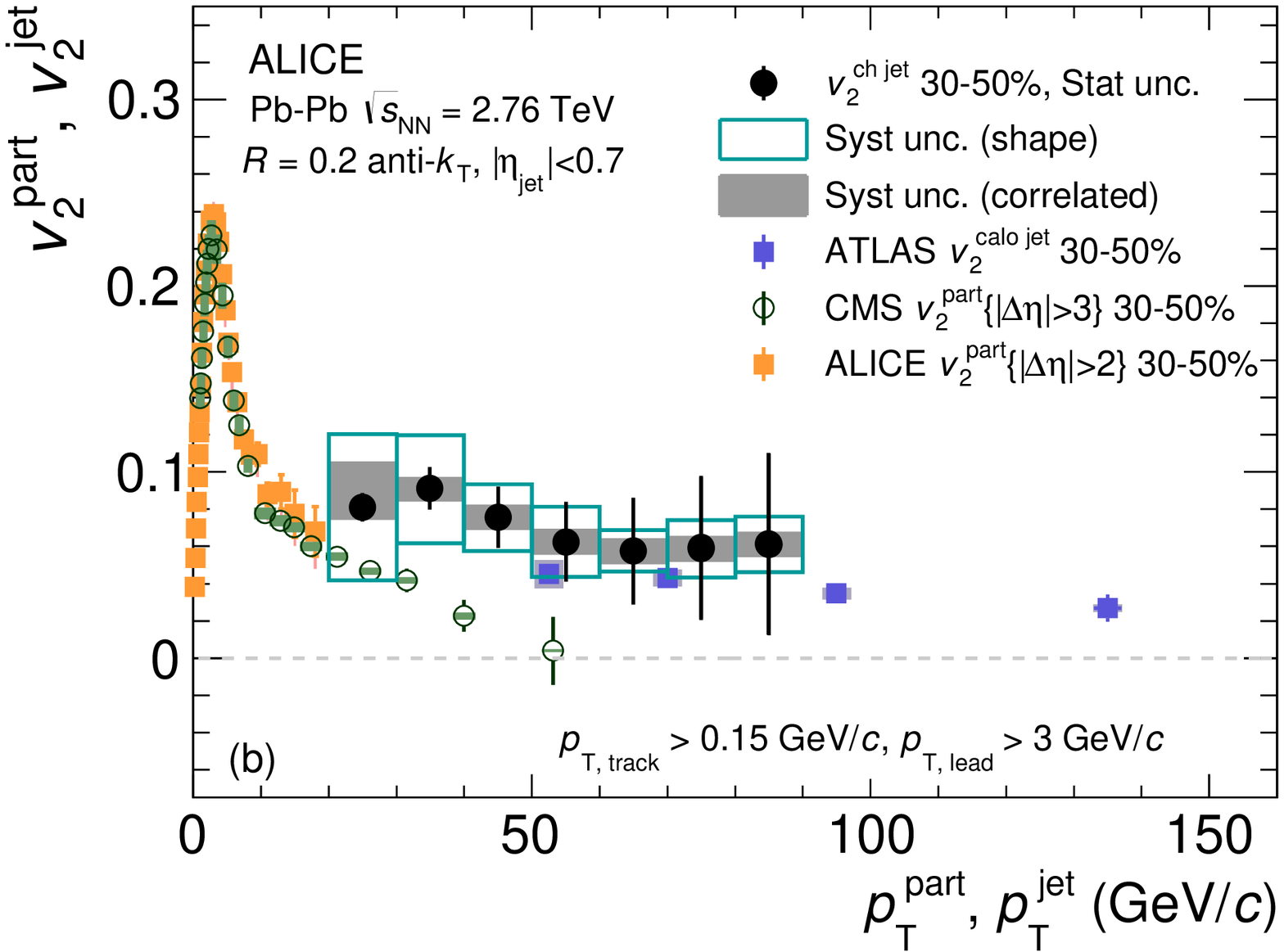}
    \caption{Elliptic flow coefficient $v_2$ of charged particles \cite{highpt, highptv2CMS} (red, green) and $R = 0.2$ full jets (comprising both charged and neutral fragments) measured within $\vert \eta \vert < 2.1$ \cite{atlas} (blue) superimposed on the results from the current analysis of $R = 0.2$ charged jets \vj{}. In all measurements, statistical errors are represented by bars and systematic uncertainties by shaded or open boxes. 
Note that the same parton \pt{} corresponds to different single particle, full jet and charged jet \pt{}. ATLAS  \vjfull and CMS $v_2$ from \cite{highptv2CMS,atlas} in 30--50 \% centrality are the weighted arithmetic means of measurements in 10\% centrality intervals using the inverse square of statistical uncertainties as weights.  } 
    \label{fig:pel}
\end{figure}

To get a better qualitative understanding of the results, the $v_2$ of single charged particles \vpart{} \cite{highpt,highptv2CMS} and the ATLAS \vjfull{} measurement \cite{atlas} are shown together with the \vj~measurement in Fig. \ref{fig:pel}. The ATLAS result is for jets with resolution parameter $R=0.2$ within $\vert \eta \vert < 2.1$ comprising both charged and neutral fragments. The event plane angle is measured by the forward calorimeter system at 3.2 $< \vert \eta \vert <$ 4.9. Jets are reconstructed by applying the anti-\kt~algorithm to calorimeter towers, after which, in an iterative procedure, a flow-modulated underlying event energy is subtracted. Each jet is required to lie within $\sqrt{\Delta \eta^2 + \Delta \varphi^2} < 0.2$ of either a calorimeter cluster of \pt~$>$~9~\GeVc~or a \pt~$>$~10~\GeVc~track jet with resolution parameter $R$ = 0.4 built from constituent tracks of \pt~$>$ 4 \GeVc{} (the full reconstruction procedure can be found in \cite{atlas, atlas2}). 
    
It is important to realize that the energy scales of the ATLAS \vjfull{} and ALICE \vj{} measurements are different (as the ALICE jets do not include neutral fragments) which complicates a direct comparison between the two measurements. The central ATLAS results are also reported in 5--10\% collision centrality. The ALICE and ATLAS measurements are in qualitative agreement, both indicating path-length-dependent parton energy loss. Given the uncertainties, the difference in the central values of the measurement is not significant.

Figure \ref{fig:pel} also shows the $v_2$ of single charged particles \vpart{} (from \cite{highpt, highptv2CMS}), which is expected to be mostly caused by in-medium energy loss at intermediate and high momenta ($\pt \gtrsim$ 5 \GeVc).  Even though a direct quantitative comparison between \vj{} and \vpart{} cannot be made as the energy scales for jets and single particles are different, the measurements can be compared qualitatively, and it can be seen that for central events, the single particle \vpart{} and \vj{} are of similar magnitude and only weakly dependent on \pt{} over a large range of \pt{} ($\approx 20 - 50$~\GeVc{}). For non-central collisions (30--50\%), the measurements of $v_2$ for single particles and jets are also in qualitative agreement in the \pt{} range where the uncertainties allow for a comparison.

\begin{figure}
    \includegraphics[width=.5\textwidth]{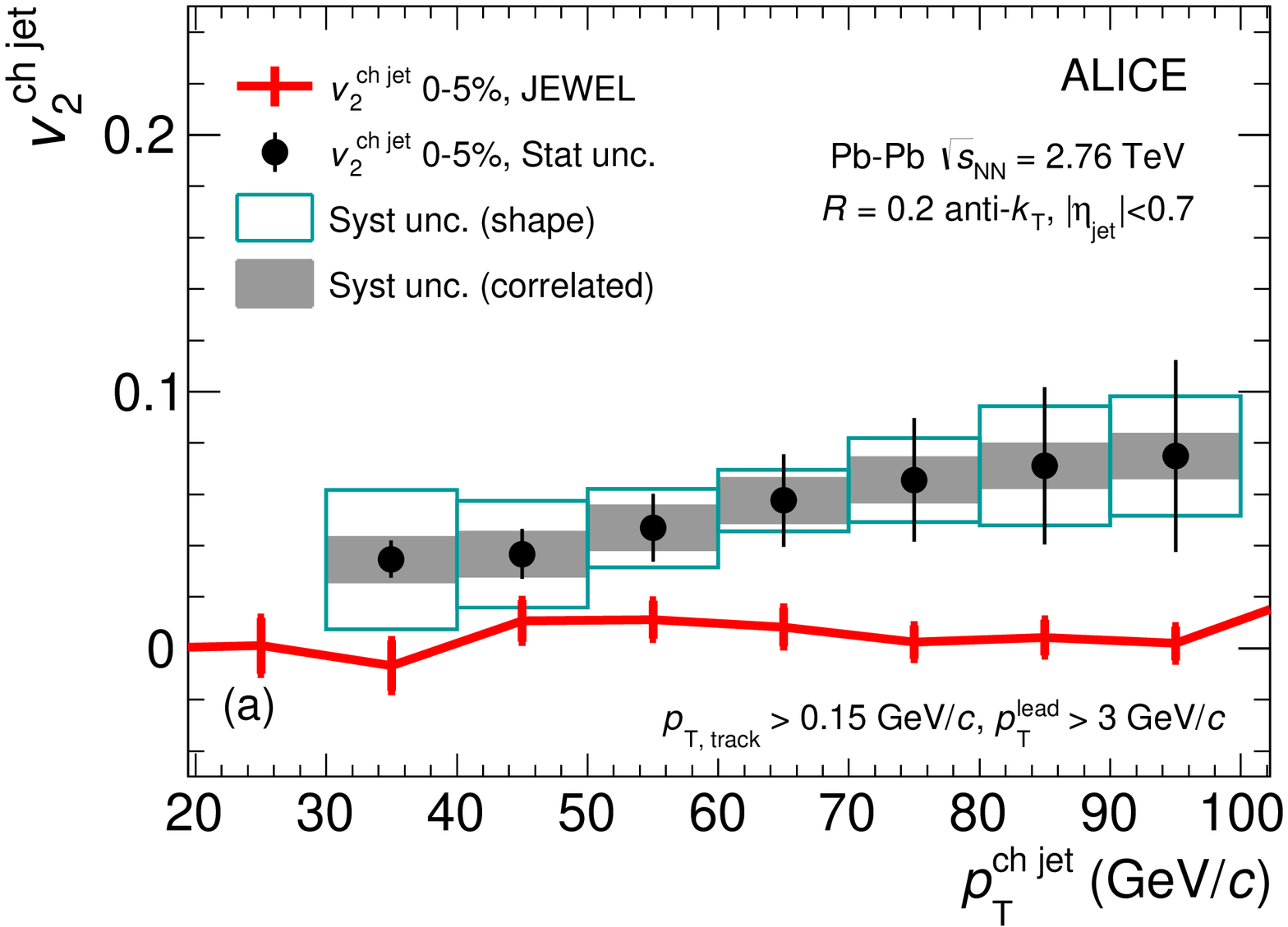}
    \includegraphics[width=.5\textwidth]{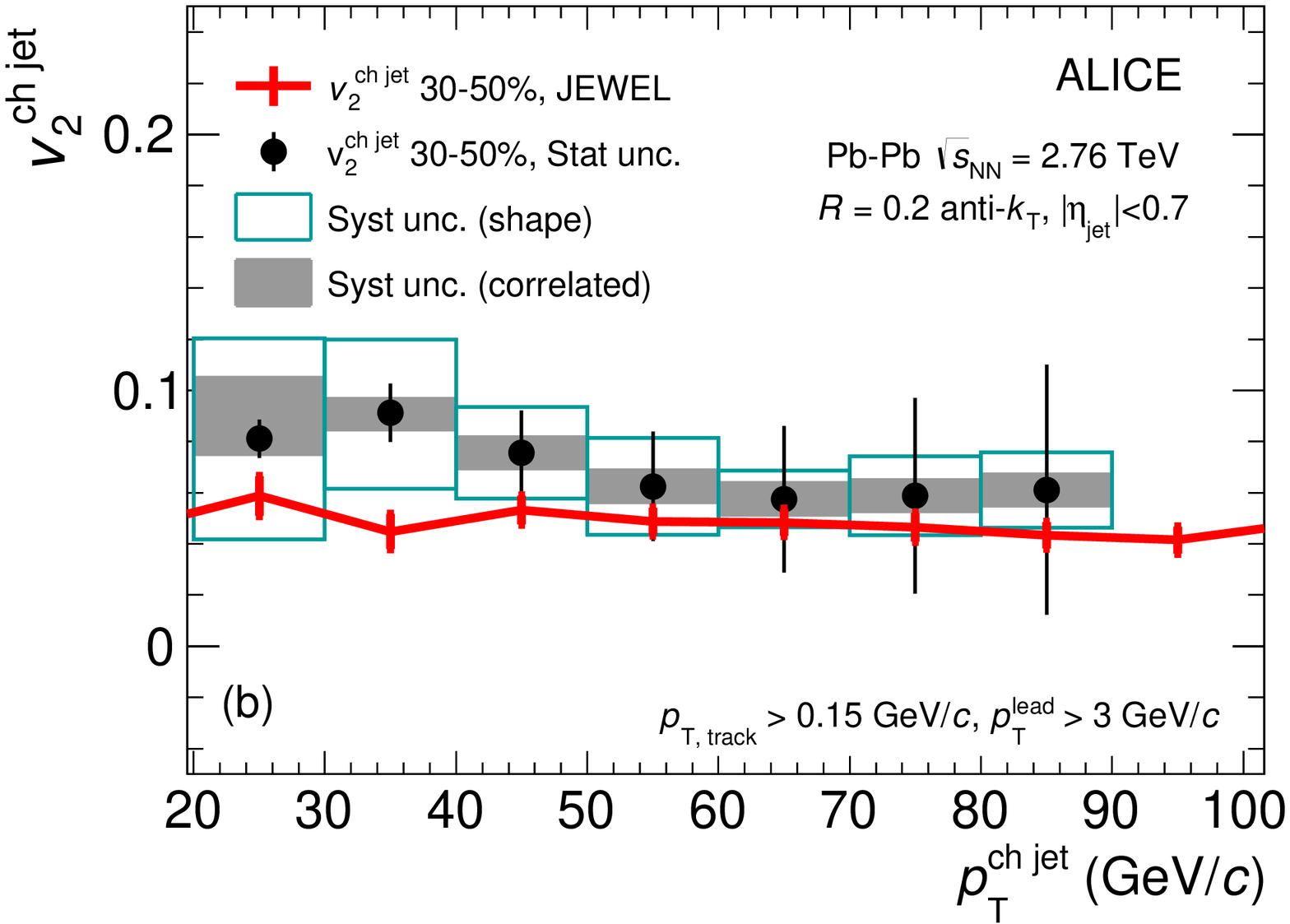}
    \caption{\vj~of $R = 0.2$ charged jets obtained from the JEWEL Monte Carlo (red) for central (a) and semi-central collisions (b) compared to data. JEWEL data points are presented with only statistical uncertainties. 
}
    \label{fig:jewel}
    \end{figure}

Figure \ref{fig:jewel} shows the \vj~of $R = 0.2$ charged jets from the JEWEL Monte Carlo \cite{jewel1,jewel2} compared to the measured \vj{}. JEWEL simulates a parton shower evolution in the presence of a dense QCD medium by generating hard scatterings according to a collision geometry from a Glauber \cite{glauber} density profile. A 1D Bjorken expansion is used to simulate the time evolution of the medium. After radiative and collisional energy loss, PYTHIA is used to hadronize the fragments to final state particles. 

The analysis on the JEWEL events is performed with the same jet definition and acceptance criteria that are used for the \vj~ analysis in data, using the symmetry plane \rptwo{} from the simulated initial geometry as \ep{}. The JEWEL Monte Carlo shows finite significant \vj{} in semi-central collisions; in central collisions \vj{} is compatible with zero. The JEWEL result for semi-central 30--50\% collisions is compatible with the measured values ($p$-value 0.4 using Eq.~\ref{a} with the JEWEL results as hypothesis $\mu_i$ and the quadratic sum of the statistical uncertainties of both datasets as $\sigma_i$ in the denominator of the first sum of Eq.~\ref{a}). 
In central JEWEL collisions \vj{} is consistent with zero, while the measured values are compatible with the JEWEL \vj{} within two standard deviations. It should also be noted that JEWEL currently uses an optical Glauber model for the initial state and therefore does not include fluctuations in the participant distribution due to the spatial configuration of nuclei in the nucleus. This simplified treatment of the overlap geometry may underestimate the \vj{} \cite{Ollitrault:1992bk,Alver:2006wh}. This comparison of \vj{} in JEWEL to experimental data complements earlier studies of the path-length-dependent parton energy loss and model predictions for the jet $R_{\mathrm{AA}}$ \cite{jets_raa}. 

\section{Conclusion}
The azimuthal anisotropy of $R = 0.2$ charged jet production, quantified as \vj, has been presented in central and semi-central collisions. Significant positive \vj~is observed in semi-central collisions, which indicates that jet suppression is sensitive to the initial geometry of the overlap region of the collision.  This observation can be used to constrain predictions on the path-length dependence of in-medium parton energy loss. In central collisions, the central values of the measurement are positive, but the uncertainties preclude drawing a strong conclusion on the magnitude of \vj{}.

The measured \vj{} for charged jets is also compared to single particle $v_2$ from ALICE and CMS and \vjfull{} from ATLAS. The measurements cannot be directly compared quantitatively since the energy scales are different, but qualitatively, the results agree and indicate a positive $v_2$ for both charged particles and jets to high \pt{} in central and semi-central collisions. This observation indicates that parton energy loss is large and that the sensitivity to the collision geometry persists up to high transverse momenta.

The JEWEL Monte Carlo predicts sizable \vj{} for semi-central collisions and very small to zero \vj{} in central events. These predictions are in good agreement with the semi-central measurement. For central collisions, the JEWEL prediction is below the measurement, but more data would be needed to reduce the uncertainties on the measurement sufficiently to constrain the model.

\newenvironment{acknowledgement}{\relax}{\relax}
\begin{acknowledgement}
    \section*{Acknowledgements}

The ALICE Collaboration would like to thank all its engineers and technicians for their invaluable contributions to the construction of the experiment and the CERN accelerator teams for the outstanding performance of the LHC complex.
The ALICE Collaboration gratefully acknowledges the resources and support provided by all Grid centres and the Worldwide LHC Computing Grid (WLCG) collaboration.
The ALICE Collaboration acknowledges the following funding agencies for their support in building and
running the ALICE detector:
State Committee of Science,  World Federation of Scientists (WFS)
and Swiss Fonds Kidagan, Armenia;
Conselho Nacional de Desenvolvimento Cient\'{\i}fico e Tecnol\'{o}gico (CNPq), Financiadora de Estudos e Projetos (FINEP),
Funda\c{c}\~{a}o de Amparo \`{a} Pesquisa do Estado de S\~{a}o Paulo (FAPESP);
National Natural Science Foundation of China (NSFC), the Chinese Ministry of Education (CMOE)
and the Ministry of Science and Technology of China (MSTC);
Ministry of Education and Youth of the Czech Republic;
Danish Natural Science Research Council, the Carlsberg Foundation and the Danish National Research Foundation;
The European Research Council under the European Community's Seventh Framework Programme;
Helsinki Institute of Physics and the Academy of Finland;
French CNRS-IN2P3, the `Region Pays de Loire', `Region Alsace', `Region Auvergne' and CEA, France;
German Bundesministerium fur Bildung, Wissenschaft, Forschung und Technologie (BMBF) and the Helmholtz Association;
General Secretariat for Research and Technology, Ministry of Development, Greece;
Hungarian Orszagos Tudomanyos Kutatasi Alappgrammok (OTKA) and National Office for Research and Technology (NKTH);
Department of Atomic Energy and Department of Science and Technology of the Government of India;
Istituto Nazionale di Fisica Nucleare (INFN) and Centro Fermi -
Museo Storico della Fisica e Centro Studi e Ricerche ``Enrico Fermi'', Italy;
MEXT Grant-in-Aid for Specially Promoted Research, Ja\-pan;
Joint Institute for Nuclear Research, Dubna;
National Research Foundation of Korea (NRF);
Consejo Nacional de Cienca y Tecnologia (CONACYT), Direccion General de Asuntos del Personal Academico(DGAPA), M\'{e}xico, Amerique Latine Formation academique - 
European Commission~(ALFA-EC) and the EPLANET Program~(European Particle Physics Latin American Network);
Stichting voor Fundamenteel Onderzoek der Materie (FOM) and the Nederlandse Organisatie voor Wetenschappelijk Onderzoek (NWO), Netherlands;
Research Council of Norway (NFR);
National Science Centre, Poland;
Ministry of National Education/Institute for Atomic Physics and National Council of Scientific Research in Higher Education~(CNCSI-UEFISCDI), Romania;
Ministry of Education and Science of Russian Federation, Russian
Academy of Sciences, Russian Federal Agency of Atomic Energy,
Russian Federal Agency for Science and Innovations and The Russian
Foundation for Basic Research;
Ministry of Education of Slovakia;
Department of Science and Technology, South Africa;
Centro de Investigaciones Energeticas, Medioambientales y Tecnologicas (CIEMAT), E-Infrastructure shared between Europe and Latin America (EELA), 
Ministerio de Econom\'{i}a y Competitividad (MINECO) of Spain, Xunta de Galicia (Conseller\'{\i}a de Educaci\'{o}n),
Centro de Aplicaciones Tecnológicas y Desarrollo Nuclear (CEA\-DEN), Cubaenerg\'{\i}a, Cuba, and IAEA (International Atomic Energy Agency);
Swedish Research Council (VR) and Knut $\&$ Alice Wallenberg
Foundation (KAW);
Ukraine Ministry of Education and Science;
United Kingdom Science and Technology Facilities Council (STFC);
The United States Department of Energy, the United States National
Science Foundation, the State of Texas, and the State of Ohio;
Ministry of Science, Education and Sports of Croatia and  Unity through Knowledge Fund, Croatia;
Council of Scientific and Industrial Research (CSIR), New Delhi, India;
Pontificia Universidad Cat\'{o}lica del Per\'{u}.
\end{acknowledgement}

\bibliographystyle{utphys}
\bibliography{references}

\newpage
\appendix
\section{The ALICE Collaboration}
\label{app:collab}



\begingroup
\small
\begin{flushleft}
J.~Adam\Irefn{org40}\And
D.~Adamov\'{a}\Irefn{org83}\And
M.M.~Aggarwal\Irefn{org87}\And
G.~Aglieri Rinella\Irefn{org36}\And
M.~Agnello\Irefn{org110}\And
N.~Agrawal\Irefn{org48}\And
Z.~Ahammed\Irefn{org132}\And
S.U.~Ahn\Irefn{org68}\And
S.~Aiola\Irefn{org136}\And
A.~Akindinov\Irefn{org58}
% ...
\renewcommand\labelenumi{\textsuperscript{\theenumi}~}

\section*{Affiliation notes}
\renewcommand\theenumi{\roman{enumi}}
\begin{Authlist}
\item \Adef{0}Deceased
\item \Adef{idp1727040}{Also at: Georgia State University, Atlanta, Georgia, United States}
\item \Adef{idp3763392}{Also at: M.V. Lomonosov Moscow State University, D.V. Skobeltsyn Institute of Nuclear, Physics, Moscow, Russia}
\end{Authlist}

\section*{Collaboration Institutes}
\renewcommand\theenumi{\arabic{enumi}~}
\begin{Authlist}

\item \Idef{org1}A.I. Alikhanyan National Science Laboratory (Yerevan Physics Institute) Foundation, Yerevan, Armenia
\item \Idef{org2}Benem\'{e}rita Universidad Aut\'{o}noma de Puebla, Puebla, Mexico
\item \Idef{org3}Bogolyubov Institute for Theoretical Physics, Kiev, Ukraine
\item \Idef{org4}Bose Institute, Department of Physics and Centre for Astroparticle Physics and Space Science (CAPSS), Kolkata, India
\item \Idef{org5}Budker Institute for Nuclear Physics, Novosibirsk, Russia
\item \Idef{org6}California Polytechnic State University, San Luis Obispo, California, United States
\item \Idef{org7}Central China Normal University, Wuhan, China
\item \Idef{org8}Centre de Calcul de l'IN2P3, Villeurbanne, France
\item \Idef{org9}Centro de Aplicaciones Tecnol\'{o}gicas y Desarrollo Nuclear (CEADEN), Havana, Cuba
\item \Idef{org10}Centro de Investigaciones Energ\'{e}ticas Medioambientales y Tecnol\'{o}gicas (CIEMAT), Madrid, Spain
\item \Idef{org11}Centro de Investigaci\'{o}n y de Estudios Avanzados (CINVESTAV), Mexico City and M\'{e}rida, Mexico
\item \Idef{org12}Centro Fermi - Museo Storico della Fisica e Centro Studi e Ricerche ``Enrico Fermi'', Rome, Italy
\item \Idef{org13}Chicago State University, Chicago, Illinois, USA
\item \Idef{org14}China Institute of Atomic Energy, Beijing, China
\item \Idef{org15}Commissariat \`{a} l'Energie Atomique, IRFU, Saclay, France
\item \Idef{org16}COMSATS Institute of Information Technology (CIIT), Islamabad, Pakistan
\item \Idef{org17}Departamento de F\'{\i}sica de Part\'{\i}culas and IGFAE, Universidad de Santiago de Compostela, Santiago de Compostela, Spain
\item \Idef{org18}Department of Physics and Technology, University of Bergen, Bergen, Norway
\item \Idef{org19}Department of Physics, Aligarh Muslim University, Aligarh, India
\item \Idef{org20}Department of Physics, Ohio State University, Columbus, Ohio, United States
\item \Idef{org21}Department of Physics, Sejong University, Seoul, South Korea
\item \Idef{org22}Department of Physics, University of Oslo, Oslo, Norway
\item \Idef{org23}Dipartimento di Elettrotecnica ed Elettronica del Politecnico, Bari, Italy
\item \Idef{org24}Dipartimento di Fisica dell'Universit\`{a} 'La Sapienza' and Sezione INFN Rome, Italy
\item \Idef{org25}Dipartimento di Fisica dell'Universit\`{a} and Sezione INFN, Cagliari, Italy
\item \Idef{org26}Dipartimento di Fisica dell'Universit\`{a} and Sezione INFN, Trieste, Italy
\item \Idef{org27}Dipartimento di Fisica dell'Universit\`{a} and Sezione INFN, Turin, Italy
\item \Idef{org28}Dipartimento di Fisica e Astronomia dell'Universit\`{a} and Sezione INFN, Bologna, Italy
\item \Idef{org29}Dipartimento di Fisica e Astronomia dell'Universit\`{a} and Sezione INFN, Catania, Italy
\item \Idef{org30}Dipartimento di Fisica e Astronomia dell'Universit\`{a} and Sezione INFN, Padova, Italy
\item \Idef{org31}Dipartimento di Fisica `E.R.~Caianiello' dell'Universit\`{a} and Gruppo Collegato INFN, Salerno, Italy
\item \Idef{org32}Dipartimento di Scienze e Innovazione Tecnologica dell'Universit\`{a} del  Piemonte Orientale and Gruppo Collegato INFN, Alessandria, Italy
\item \Idef{org33}Dipartimento Interateneo di Fisica `M.~Merlin' and Sezione INFN, Bari, Italy
\item \Idef{org34}Division of Experimental High Energy Physics, University of Lund, Lund, Sweden
\item \Idef{org35}Eberhard Karls Universit\"{a}t T\"{u}bingen, T\"{u}bingen, Germany
\item \Idef{org36}European Organization for Nuclear Research (CERN), Geneva, Switzerland
\item \Idef{org37}Excellence Cluster Universe, Technische Universit\"{a}t M\"{u}nchen, Munich, Germany
\item \Idef{org38}Faculty of Engineering, Bergen University College, Bergen, Norway
\item \Idef{org39}Faculty of Mathematics, Physics and Informatics, Comenius University, Bratislava, Slovakia
\item \Idef{org40}Faculty of Nuclear Sciences and Physical Engineering, Czech Technical University in Prague, Prague, Czech Republic
\item \Idef{org41}Faculty of Science, P.J.~\v{S}af\'{a}rik University, Ko\v{s}ice, Slovakia
\item \Idef{org42}Faculty of Technology, Buskerud and Vestfold University College, Vestfold, Norway
\item \Idef{org43}Frankfurt Institute for Advanced Studies, Johann Wolfgang Goethe-Universit\"{a}t Frankfurt, Frankfurt, Germany
\item \Idef{org44}Gangneung-Wonju National University, Gangneung, South Korea
\item \Idef{org45}Gauhati University, Department of Physics, Guwahati, India
\item \Idef{org46}Helsinki Institute of Physics (HIP), Helsinki, Finland
\item \Idef{org47}Hiroshima University, Hiroshima, Japan
\item \Idef{org48}Indian Institute of Technology Bombay (IIT), Mumbai, India
\item \Idef{org49}Indian Institute of Technology Indore, Indore (IITI), India
\item \Idef{org50}Inha University, Incheon, South Korea
\item \Idef{org51}Institut de Physique Nucl\'eaire d'Orsay (IPNO), Universit\'e Paris-Sud, CNRS-IN2P3, Orsay, France
\item \Idef{org52}Institut f\"{u}r Informatik, Johann Wolfgang Goethe-Universit\"{a}t Frankfurt, Frankfurt, Germany
\item \Idef{org53}Institut f\"{u}r Kernphysik, Johann Wolfgang Goethe-Universit\"{a}t Frankfurt, Frankfurt, Germany
\item \Idef{org54}Institut f\"{u}r Kernphysik, Westf\"{a}lische Wilhelms-Universit\"{a}t M\"{u}nster, M\"{u}nster, Germany
\item \Idef{org55}Institut Pluridisciplinaire Hubert Curien (IPHC), Universit\'{e} de Strasbourg, CNRS-IN2P3, Strasbourg, France
\item \Idef{org56}Institute for Nuclear Research, Academy of Sciences, Moscow, Russia
\item \Idef{org57}Institute for Subatomic Physics of Utrecht University, Utrecht, Netherlands
\item \Idef{org58}Institute for Theoretical and Experimental Physics, Moscow, Russia
\item \Idef{org59}Institute of Experimental Physics, Slovak Academy of Sciences, Ko\v{s}ice, Slovakia
\item \Idef{org60}Institute of Physics, Academy of Sciences of the Czech Republic, Prague, Czech Republic
\item \Idef{org61}Institute of Physics, Bhubaneswar, India
\item \Idef{org62}Institute of Space Science (ISS), Bucharest, Romania
\item \Idef{org63}Instituto de Ciencias Nucleares, Universidad Nacional Aut\'{o}noma de M\'{e}xico, Mexico City, Mexico
\item \Idef{org64}Instituto de F\'{\i}sica, Universidad Nacional Aut\'{o}noma de M\'{e}xico, Mexico City, Mexico
\item \Idef{org65}iThemba LABS, National Research Foundation, Somerset West, South Africa
\item \Idef{org66}Joint Institute for Nuclear Research (JINR), Dubna, Russia
\item \Idef{org67}Konkuk University, Seoul, South Korea
\item \Idef{org68}Korea Institute of Science and Technology Information, Daejeon, South Korea
\item \Idef{org69}KTO Karatay University, Konya, Turkey
\item \Idef{org70}Laboratoire de Physique Corpusculaire (LPC), Clermont Universit\'{e}, Universit\'{e} Blaise Pascal, CNRS--IN2P3, Clermont-Ferrand, France
\item \Idef{org71}Laboratoire de Physique Subatomique et de Cosmologie, Universit\'{e} Grenoble-Alpes, CNRS-IN2P3, Grenoble, France
\item \Idef{org72}Laboratori Nazionali di Frascati, INFN, Frascati, Italy
\item \Idef{org73}Laboratori Nazionali di Legnaro, INFN, Legnaro, Italy
\item \Idef{org74}Lawrence Berkeley National Laboratory, Berkeley, California, United States
\item \Idef{org75}Moscow Engineering Physics Institute, Moscow, Russia
\item \Idef{org76}Nagasaki Institute of Applied Science, Nagasaki, Japan
\item \Idef{org77}National Centre for Nuclear Studies, Warsaw, Poland
\item \Idef{org78}National Institute for Physics and Nuclear Engineering, Bucharest, Romania
\item \Idef{org79}National Institute of Science Education and Research, Bhubaneswar, India
\item \Idef{org80}Niels Bohr Institute, University of Copenhagen, Copenhagen, Denmark
\item \Idef{org81}Nikhef, Nationaal instituut voor subatomaire fysica, Amsterdam, Netherlands
\item \Idef{org82}Nuclear Physics Group, STFC Daresbury Laboratory, Daresbury, United Kingdom
\item \Idef{org83}Nuclear Physics Institute, Academy of Sciences of the Czech Republic, \v{R}e\v{z} u Prahy, Czech Republic
\item \Idef{org84}Oak Ridge National Laboratory, Oak Ridge, Tennessee, United States
\item \Idef{org85}Petersburg Nuclear Physics Institute, Gatchina, Russia
\item \Idef{org86}Physics Department, Creighton University, Omaha, Nebraska, United States
\item \Idef{org87}Physics Department, Panjab University, Chandigarh, India
\item \Idef{org88}Physics Department, University of Athens, Athens, Greece
\item \Idef{org89}Physics Department, University of Cape Town, Cape Town, South Africa
\item \Idef{org90}Physics Department, University of Jammu, Jammu, India
\item \Idef{org91}Physics Department, University of Rajasthan, Jaipur, India
\item \Idef{org92}Physik Department, Technische Universit\"{a}t M\"{u}nchen, Munich, Germany
\item \Idef{org93}Physikalisches Institut, Ruprecht-Karls-Universit\"{a}t Heidelberg, Heidelberg, Germany
\item \Idef{org94}Purdue University, West Lafayette, Indiana, United States
\item \Idef{org95}Pusan National University, Pusan, South Korea
\item \Idef{org96}Research Division and ExtreMe Matter Institute EMMI, GSI Helmholtzzentrum f\"ur Schwerionenforschung, Darmstadt, Germany
\item \Idef{org97}Rudjer Bo\v{s}kovi\'{c} Institute, Zagreb, Croatia
\item \Idef{org98}Russian Federal Nuclear Center (VNIIEF), Sarov, Russia
\item \Idef{org99}Russian Research Centre Kurchatov Institute, Moscow, Russia
\item \Idef{org100}Saha Institute of Nuclear Physics, Kolkata, India
\item \Idef{org101}School of Physics and Astronomy, University of Birmingham, Birmingham, United Kingdom
\item \Idef{org102}Secci\'{o}n F\'{\i}sica, Departamento de Ciencias, Pontificia Universidad Cat\'{o}lica del Per\'{u}, Lima, Peru
\item \Idef{org103}Sezione INFN, Bari, Italy
\item \Idef{org104}Sezione INFN, Bologna, Italy
\item \Idef{org105}Sezione INFN, Cagliari, Italy
\item \Idef{org106}Sezione INFN, Catania, Italy
\item \Idef{org107}Sezione INFN, Padova, Italy
\item \Idef{org108}Sezione INFN, Rome, Italy
\item \Idef{org109}Sezione INFN, Trieste, Italy
\item \Idef{org110}Sezione INFN, Turin, Italy
\item \Idef{org111}SSC IHEP of NRC Kurchatov institute, Protvino, Russia
\item \Idef{org112}Stefan Meyer Institut f\"{u}r Subatomare Physik (SMI), Vienna, Austria
\item \Idef{org113}SUBATECH, Ecole des Mines de Nantes, Universit\'{e} de Nantes, CNRS-IN2P3, Nantes, France
\item \Idef{org114}Suranaree University of Technology, Nakhon Ratchasima, Thailand
\item \Idef{org115}Technical University of Ko\v{s}ice, Ko\v{s}ice, Slovakia
\item \Idef{org116}Technical University of Split FESB, Split, Croatia
\item \Idef{org117}The Henryk Niewodniczanski Institute of Nuclear Physics, Polish Academy of Sciences, Cracow, Poland
\item \Idef{org118}The University of Texas at Austin, Physics Department, Austin, Texas, USA
\item \Idef{org119}Universidad Aut\'{o}noma de Sinaloa, Culiac\'{a}n, Mexico
\item \Idef{org120}Universidade de S\~{a}o Paulo (USP), S\~{a}o Paulo, Brazil
\item \Idef{org121}Universidade Estadual de Campinas (UNICAMP), Campinas, Brazil
\item \Idef{org122}University of Houston, Houston, Texas, United States
\item \Idef{org123}University of Jyv\"{a}skyl\"{a}, Jyv\"{a}skyl\"{a}, Finland
\item \Idef{org124}University of Liverpool, Liverpool, United Kingdom
\item \Idef{org125}University of Tennessee, Knoxville, Tennessee, United States
\item \Idef{org126}University of the Witwatersrand, Johannesburg, South Africa
\item \Idef{org127}University of Tokyo, Tokyo, Japan
\item \Idef{org128}University of Tsukuba, Tsukuba, Japan
\item \Idef{org129}University of Zagreb, Zagreb, Croatia
\item \Idef{org130}Universit\'{e} de Lyon, Universit\'{e} Lyon 1, CNRS/IN2P3, IPN-Lyon, Villeurbanne, France
\item \Idef{org131}V.~Fock Institute for Physics, St. Petersburg State University, St. Petersburg, Russia
\item \Idef{org132}Variable Energy Cyclotron Centre, Kolkata, India
\item \Idef{org133}Warsaw University of Technology, Warsaw, Poland
\item \Idef{org134}Wayne State University, Detroit, Michigan, United States
\item \Idef{org135}Wigner Research Centre for Physics, Hungarian Academy of Sciences, Budapest, Hungary
\item \Idef{org136}Yale University, New Haven, Connecticut, United States
\item \Idef{org137}Yonsei University, Seoul, South Korea
\item \Idef{org138}Zentrum f\"{u}r Technologietransfer und Telekommunikation (ZTT), Fachhochschule Worms, Worms, Germany
\end{Authlist}
\endgroup

\end{document}